\documentclass[manuscript]{emulateapj}
\usepackage[pdftex]{hyperref}

\newcommand\plotme[1]{\centering\includegraphics[width=8.5cm]{#1}}

\newcommand\kms{{\rm \,km\,s^{-1}}}

\newcommand\beq{\begin{equation}}
\newcommand\eeq{\end{equation}}

\newcommand\vc{\rm{v_{\rm LSR}}}

\newcommand\vi{\rm{v_i}}
\newcommand\vj{\rm{v_j}}
\newcommand\vphi{\rm{v_\phi}}
\newcommand\vz{\rm{v_z}}
\newcommand\vR{\rm{v_R}}

\newcommand\sm{{\rm M}_\odot}
\newcommand\masyr{${\rm mas\,yr}^{-1}$}
\newcommand\feh{[{\rm Fe/H}]}
\newcommand\vrsq{\sigma_{\rm R}^2}

\newcommand\vrvphi{\sigma_{\rm{R\phi}}}
\newcommand\dispvr{\sigma_{\rm{R}}}
\newcommand\dispvphi{\sigma_{\rm{\phi}}}
\newcommand\dispvz{\sigma_{\rm z}}
\newcommand\covrz{\sigma_{\rm{Rz}}}
\newcommand\covrphi{\sigma_{\rm{R\phi}}}

\shorttitle{Slicing and dicing the Milky Way disk}
\shortauthors{Smith, Whiteoak \& Evans}
\slugcomment{Submitted to The Astrophysical Journal}

\begin{document}
\setcounter{table}{1}

\title{Slicing and dicing the Milky Way disk in the Sloan Digital Sky Survey}

\author{Martin C. Smith\altaffilmark{1,2},
S. Hannah Whiteoak\altaffilmark{3},
and N. W. Evans\altaffilmark{3}}
\altaffiltext{1}{Kavli Institute for Astronomy and Astrophysics, Peking
  University, Beijing 100871, China; \mbox{msmith@pku.edu.cn}}
\altaffiltext{2}{National Astronomical Observatories, Chinese Academy of
  Sciences, Beijing 100012, China}
\altaffiltext{3}{Institute of Astronomy, University of Cambridge, Madingley
  Road, Cambridge CB3 0HA, UK}

\begin{abstract}
We use the Stripe 82 proper motion catalogue of \citet{Br08} to study
the kinematics of Galactic disk stars in the solar neighborhood. We
select samples of dwarf stars with reliable spectra and proper
motions. They have cylindrical polar radius between $7 \le \rm{R} \le
9$ kpc, heights from the Galactic plane satisfying $|z| \le 2$ kpc and
span a range of metallicities $-1.5 \le \feh \le 0$. We develop a
method for calculating and correcting for the halo contamination in
our sample using the distribution of rotational velocities.
%  For all but our most metal-poor samples, the fraction of halo stars
%  is less than $5$ per cent.
Two Gaussians representing disk and halo populations are used to fit
the radial ($\vR$) and vertical ($\vz$) velocity distributions
via maximum likelihood methods. For the azimuthal
velocities ($\vphi$) the same technique is used, except that a skewed
non-Gaussian functional form now represents the disk velocity
distribution. This enables us to compute the dispersions $\dispvr,
\dispvz, \dispvphi$ and cross-terms (the tilt $\covrz$ and the vertex
deviation $\covrphi$) of the velocity ellipsoid as a function of
height and metallicity. We also investigate the rotation lag of the
disk, finding that the more metal-poor stars rotate significantly
slower than the metal-rich stars.  These samples provide important
constraints on heating mechanisms in the Galactic disk and can be used
for a variety of applications. We present one such application,
employing the Jeans equations to provide a simple model of the
potential close to the disk. Our model is in excellent agreement with
others in the literature and provides an indication the disk, rather
than the halo, dominates the circular speed at the solar
neighborhood. We obtain a surface mass density within 1.1 kpc
of around $66 \: \sm {\rm pc}^{-2}$ and estimate a local halo density of
$0.015\:\sm {\rm pc}^{-3} = 0.57\:{\rm GeV\;cm^{-3}}$.
\end{abstract}


\section{Introduction}

In the simplest model of a spiral galaxy, stars in the disk librate
about circular orbits in the equatorial plane (``the epicyclic
approximation'', \citealt*{BT}). This picture is complicated
by astrophysical processes: such as bar instabilities, satellite
accretion, scattering by spiral arms or collisions with molecular
clouds.  Surveys which gather kinematic data allow astronomers to
analyze the properties of large samples of stars and hence probe the
signatures left by heating processes in their net motions and
dispersions. For example, the {\it Hipparcos} satellite provided
precise parallaxes and proper motions for nearby stars, which were
used to provide the velocity dispersion as a function of stellar type
in the immediate solar neighborhood \citep{DB}.

In their 1983 study of the density distribution of stars up to 4 kpc
below the plane, Gilmore \& Reid argue that the disk is in fact
composed of two distinct components: a thin disk with approximate
scale height $\sim 300$ pc, and an older, more metal-poor thick
disk. Further studies (see \citealt{Reid93} for a review) have supported
this claim, although the thick disk scale height remains poorly
constrained \citep[eg.][]{Juric08,deJ10} and the question of how the
two-component structure arose is still open.  Suggested scenarios
include resonant trapping, which converts planar orbits into inclined
ones \citep{SridharTouma}, or the dissipational collapse of gas
\citep{Br04}, or the accretion of a satellite galaxy, in which case
the thick disk may form from satellite debris, or as a result of
heating of the thin disk by the collision \citep[eg.][]{Vi08}. In
recent years a lot of attention has focussed on the formation of thick
disks through radial migration, where stars trapped in co-rotation
resonances of the spiral arms are transported
through the disk \citep{Se02,Ro08,Sc09} or through similar mechanisms
due to the Galactic bar \citep{Mi10}. Detailed tests of the chemical
properties of the Milky Way disk are also becoming possible, for
example using large samples of stars with alpha-element abundances to
characterize the origins of the different components
\citep[e.g.][]{Bo11,Le11,Na11,Ru10}.

Much can also be learnt about heating processes through studying the
size and shape of the stellar velocity ellipsoid, or equivalently the
stellar velocity dispersion tensor. The ratio of radial to vertical
velocity dispersion $\dispvr/\dispvz$, as well as the dependence of
the dispersion on the age of a stellar population, can be compared
with analytical and numerical predictions for various models of disk
heating, to see which mechanisms dominate in our galaxy~\citep{Fu87}.
The ratio of tangential to radial velocity dispersion
$\dispvphi/\dispvr$ is predicted to be $1/\sqrt{2}$ from epicylic
theory for a galaxy with a flat rotation curve, though it has long
been known to be less than this for the Milky Way \citep{Kerr86,
  EC93}.  The covariances $\covrz$ and $\covrphi$ are also of
interest; non-zero $\covrphi$ implies that the Galactic potential is
non-axisymmetric \citep{BT}.  The quantity $\covrz$, or more
specifically, its vertical gradient, is important for the calculation
of Galactic parameters, including the local surface mass density of
the disk \citep{Ku89} and the asymmetric drift relation~\citep{DB}.

Large-scale photometric and kinematic surveys are making measurements
of the components of the velocity dispersion tensor possible for
populations in the Milky Way Galaxy. For example, using data from the
Radial Velocity Experiment (RAVE), \citet{Siebert08} measured
the magnitude and orientation of the velocity dispersion tensor for
red clump stars between 500 pc and 1500 pc below the Galactic
Plane. They reckoned the tilt to be $7.3 \pm 1.8^\circ$, which is consistent
with alignment in spherical polar coordinates. \citeauthor{Siebert08}
compare this value to computed inclinations for two mass models of the
Milky Way. The measurement is reproducible with a short scalelength of
the stellar disk ($\approx 2$ kpc) if the dark halo is oblate or with
a long scalelength ($\approx 3$ kpc) if the dark halo is spherical or
prolate. A similar study has been published subsequently by
\citet{Ca11}, also using radial velocities from the RAVE survey,
finding a tilt of $8.6\pm1.8^\circ$ for red clump stars at a distance
of 700 to 2000 pc from the plane.

The kinematics properties of the Galactic disk can also be used to
probe directly its gravitational potential, as was carried out in the
classical work of \citet{Ku91}. Using a sample of around 500 K dwarfs,
they were able to estimate that the surface mass density within 1.1
kpc was $71\pm6\: \sm {\rm pc}^{-2}$.  Subsequent work has supported
this measurement, for example, \citet{Ho04}.  An inventory of the
surface density of baryonic material in the solar neighborhood
indicates that around a third of this mass is likely non-baryonic
\citep[e.g.][]{BE}.  Such techniques can also be used to determine the
local mass density and hence make predictions for the local dark
matter density \citep[e.g.][]{Ho00,Ga11}, which is a vital ingredient
in predictions for the direct detection of dark matter \citep[][and
references therein]{Be05}.

Clearly obtaining unbiased
determinations of the kinematic properties of the disk, including
their gradients, is of great importance for understanding its
structure and evolution.
Here, we study the kinematics of disk stars using data from the
recently constructed Sloan Digital Sky Survey (SDSS) Stripe 82
catalogue~\citep{Br08}. This covers an equatorial stripe of area $250$
deg$^2$, which has been repeatedly imaged by the SDSS since 1998
primarily with the aim of supernova discovery. The catalogue contains
almost 4 million ``light-motion curves'' and is complete down to a
magnitude of $r \approx 21.5$.  It reaches almost 2 mag fainter than
the SDSS/USNO-B catalogue \citep{Mu04}, making it the deepest
large-area photometric and astrometric catalogue available. We have
already exploited the Stripe 82 catalogue to measure the components of
the velocity dispersion tensor for a sample of halo stars
\citep[e.g.][]{Sm09,Sm09b}. Here, we provide a complementary study for
disk populations.

In Section \ref{sec:sample}, we describe the selection cuts that we
apply to the Stripe 82 catalogue to generate our sample of dwarf
stars.  The procedure for splitting the sample into disk and halo
components is described in Section \ref{sec:analysis}, as well as a
discussion of the effects of measurement errors. Our results are
presented and discussed in Section \ref{sec:properties}.

\section{Sample construction}
\label{sec:sample}

We use data from the 7th public data release of the Sloan Digital Sky
Survey \citep[SDSS;][]{Ab09,Ya09}.  We restrict attention to stars in
the spectroscopic part of the survey, so as to have estimates of
velocities, metallicities and surface gravities.  The full details
describing the analysis of these data, in particular the reliability
of the derived velocities and stellar parameters, can be found in
\citet{Le08} and references therein.

Although SDSS data cover a huge part of the sky, mainly in the
Northern Galactic cap, we also restrict ourselves to data in Stripe
82, so as to exploit the catalogue of high-precision photometry and
proper-motions \citep{Br08}. This stripe covers around 250 square
degrees in region $\alpha = 20.7$ hr to 3.3 hr and
$\left|\delta\right| < 1.26^\circ$. This corresponds to the South
Galactic cap, with $l \sim 50$ to 190 degrees and $b \sim -25$ to
$-60$ degrees.

From this data, we select stars with reliable spectra (i.e. labelled
in the catalogue by the flag 'nnnn') and errors in $\log{g}$ and $\feh$
of less than 0.5 dex and in radial velocity of less than 20 $\kms$. We
require proper motion errors of less than 4 \masyr.
%% Stars with valid spectral measurements are labelled in the catalogue
%% by the flag 'nnnn'.  This entry indicates that the spectrum has passed
%% the tests carried out by \citet{Le08}.  These tests check that the
%% star is main-sequence and that the spectrum is not noisy, does not
%% contain unexpected emission lines, or give a (g-r) value very
%% different to the value determined by photometry for that star.
We also place a cut in $\log{g}$ in order to remove giant stars from
the sample, which we place at $\log{g} = 3.5$.  To investigate the
effects of giant contamination, we repeat our analysis of the sample
properties using the stricter selection criterion $\log{g} \ge 4$.
This stricter cut excludes only $\sim 1$ per cent of our sample. We
checked that this made no qualitative differences to the trends
observed.

The errors quoted in the SDSS stellar parameter pipeline are internal
errors. Comparison with observations made using high-resolution
spectroscopy reveals that there are also external errors of 2.4
$\kms$, 0.11 dex and 0.21 dex present in the radial velocities,
metallicities and surface gravities, respectively \citep{Prieto08}.
We add the internal and external errors in quadrature to obtain the
overall error in each measurement.

We correct all magnitudes for the effects of extinction by
interstellar dust, using the maps of \citet{Schlegel}. We also place a
cut in color of $0.3 \le (g-i) \le 4.0$. This is required so that we
can obtain reliable photometric distances, which we estimate using the
relation of \citet[][equations A1-A5]{Iv08}. We make one minor
modification to the relation of \citet{Iv08}, namely we do not apply
the turn-off correction given by equation (A6) of their paper. Their
correction is based on the metal-poor globular cluster M13, which may
not be suitable for disk stars. In Appendix \ref{app:turn-off} we use
stellar models to show that a better choice may be to simply neglect
this correction term.  Note that although \citet{Iv08} use photometric
estimates for the metallicities in their paper, we are able to use the
more reliable spectroscopic metallicities for our sample.
It is worth pointing out that although the \citet{Iv08} paper deals
with the analysis of photometric metallicities, their parallax
relation is derived from star clusters; as a consequence our work will
not be affected by any systematic offsets between spectroscopic and
photometric metallicities (should any be present).
The median error on our distances is $10.7$ per cent.

Once we have an estimate for the distance, we can determine the full
three-dimensional positions and velocities. We express these in terms
of Galactocentric cylindrical coordinates, $\rm{(R,\phi,z)}$, where
$\rm{\phi}$ is defined as increasing in the direction opposite to the
solar rotation, and $z$ is positive towards the North Galactic Pole
(NGP).  We calculate corresponding velocities $(\vR,\vphi,\vz)$ for
each star, and correct for the solar motion \citep{DB} and the motion
of the local standard of rest (LSR) so that our stellar velocities are
relative to the Galactic frame, ie. $\vc = -220 \kms$.

The errors on the positions and velocities are calculated using a
Monte Carlo method. This is done by taking the uncertainties on all of
the observed quantities for a given star (proper motion, radial
velocity, magnitudes and metallicity) and randomly sampling from these
assuming the errors are Gaussian; for each realisation we then
calculate a distance (including uncertainties in the photometric
parallax relation) and hence a velocity. We repeat this sampling 1000
times and from the resulting distributions of positions and velocities
calculate our error matrix for a given star.
The median values of the errors $\delta\vR$,
$\delta\vphi$ and $\delta\vz$ are 27.2, 27.3 and 16.9 $\kms$,
respectively. As our line of sight is approximately aligned with the
South Galactic Pole, $\vz$ is mostly determined by radial velocity
measurements, whereas $\vR$ and $\vphi$ are mostly dependent on proper
motions.  The measurement errors in the proper motions are much larger
than in the radial velocities, hence $\delta \vR$ and $\delta \vphi$
are larger than $\delta \vz$. We also calculate the correlations in
the errors (e.g. $\delta\vR\vz$), which are important for the
covariances.

\section{Algorithms for Disk Kinematics}
\label{sec:analysis}

Here, we describe the techniques used to obtain the kinematic
properties of the disk. We aim to study the properties of our sample
as a function of height above or below the Galactic plane
$z$. However, many of these properties depend strongly on metallicity,
so to allow for the influence of vertical metallicity gradients in the
disk, we split the sample into ranges in metallicity and study trends
in each range separately. This is also crucial because the SDSS
spectroscopic selection function can introduce biases which are
difficult to model. Therefore, by binning our data in metallicity
ranges, we are only making the assumption that the metallicity
distribution {\it within a bin} is representative of the true
metallicity distribution. Although this will not be exactly the case
due to variations within a bin, both in terms of metallicity and age,
it should remove the main source of systematic bias in the analysis.

The range of our data in cylindrical polar radius $R$ is too narrow
for us to determine trends in the radial direction with any
confidence. Clearly, the further we go from the plane, the larger the
range in $R$ which is covered, and so in order to keep our data to a
specific range in $R$ we restrict ourselves to $7 \le \rm{R} \le
9$. Therefore, we split the data into three ranges in metallicity ($-1.5
\le \feh \le -0.8, -0.8 \le \feh \le -0.5$ and $-0.5 \le \feh$), and
then for each metallicity bin we further divide the data into four
ranges in z out to a maximum distance of 2 kpc. A total of 7280
stars match these criteria. The stars are equally divided between the
four distance bins, resulting in in around $500$ to $800$ stars per
bin (see Table \ref{tab:kinematics}).

\subsection{The Stellar Halo Contribution}
\label{sec:split}

In order to carry out an unbiased study of the kinematics of the
disk, we need a method to model the contamination from the halo
stars in our sample. To do this, we make the
assumption that all counter-rotating ($\vphi > 0$) stars belong to the
halo, i.e., the number of halo stars is simply twice the number of
counter-rotating stars.  This provides an estimate for the level of
halo contamination in each bin, which allows us to make corrections
when calculating our kinematic properties. Some authors have argued
that the local stellar halo is rotating \citep[e.g.][]{Ca07}, although
this is in conflict with later studies \citep[e.g.][]{Iv08,Sm09} which
show that it is consistent with little or zero rotation.\footnote{Our
results have been found using the assumption that the halo is not
rotating. However, for comparison purposes, we investigate the effect
that a rotating halo would have on our results \citep[following the
results of][]{Ca07}.  This has a tendency to slightly increase our
estimated halo fractions, but the effect on our results are small and
our conclusions remain unchanged. Note also that the halo rotation is
entirely degenerate with the assumed value for the local standard of
rest, which we have taken to be 220 $\kms$ \citep[see, for
example,][]{De11}.}

The resulting halo fractions are listed in
Table~\ref{tab:kinematics}, from which it can be seen that in
general halo contamination is small. For all except the most
metal-poor bins, the fraction of halo stars is less than 5 per cent
and usually no more than 1 per cent. However, the lowest range in
metallicity, which contains the largest fraction of halo stars, may be
susceptible to halo contamination if the subtraction is imperfect.  As
expected, the number of halo stars increases as we move further from
the plane. It should be noted that due to the spectroscopic selection
function of the SDSS survey, we do not expect these values to be
representative of a volume-limited sample.

%Therefore, we do not draw any scientific conclusions from this
%figure beyond the effect that this level of halo contamination will
%have on the following kinematic analysis.

%\begin{figure}
%\plotone{f_H}
%\plotone{halo_fraction.png}
%\caption{The fraction of halo stars in each of our bins. The black,
%  blue and cyan points correspond to metallicity ranges $-1.5 \le \feh
%  \le -0.8$, $-0.8 \le \feh \le -0.5$ and $-0.5 \le \feh$,
%  respectively.  Note that due to the spectroscopic selection function
%  of the SDSS survey we do not necessarily expect these values to be
%  representative of a volume limited sample, i.e. one should not drawn
%  any scientific conclusions from this figure beyond the effect that
%  this level of halo contamination will have on the following
%  kinematic analysis.  }
%\label{fig:fraction_z}
%\end{figure}

\subsection{Velocity Dispersions}
\label{sec:method_disp}

%Fig. \ref{fig:vrvzfit} shows the $\vR$ and $\vz$ distributions for
%each bin in z for stars in the metallicity range $-1.5 \le \feh \le
%-0.8$.

We use a maximum likelihood method to fit each $\vR$ and $\vz$
distribution as the sum of the disk and halo distributions, which are
modelled using Gaussians. Although these velocity distributions are
not expected to be exactly Gaussian, a maximum likelihood technique
provides more robust results than simply calculating the sample
variance.  The relative normalization is the halo fraction as
calculated in Section \ref{sec:split}, and the velocity dispersions
for the halo are fixed at $\dispvr = 138.2\kms$ and $\dispvz =
89.3\kms$ \citep{Sm09} with no bulk motion. Note that in the
calculation of the maximum likelihood, we take into account the errors
on the individual velocities, as estimated in Section
\ref{sec:sample}.

We also use a maximum likelihood method to fit the $\vphi$
distribution. However, it is well known that the distribution of
$\vphi$ for the disk is highly skewed and non-Gaussian
\citep[e.g.][]{St27,EC93,Cu94,BM}. Therefore, we require an asymmetric
model to fit this distribution and adopt the functional form
described in equation (15) of \citet{Cu94}, namely the distribution
function for $\vphi$ is given by,
\begin{eqnarray}
\label{eq:cu94}
f(\vphi) & \propto &
{\rm exp}
\left(
-\frac{\vR^2}{2\sigma_{\rm d}^2}{\rm e}^{y\vphi/{\rm v_c}}
\right)\\
&&\times\:
{\rm exp}
\left\{
\frac{1}{2\sigma_{\rm d}^2}
\left[
\vphi^2-{\rm v_c}^2 + 2{\rm v_c}^2 {\rm ln}\left(\frac{{\rm v_c}}{\vphi}\right)
\right]
{\rm e}^{y\vphi/{\rm v_c}}
\right\},\nonumber
\end{eqnarray}
where $y = 8\:{\rm kpc} / {\rm R_d}$.

This results in a prediction for the distribution of $\vphi$ as a
function of two free parameters (${\rm v_c}$ and ${\rm R_d}$).
To remove the dependence on $\vR$ in this equation we marginalise
over the known distribution of $\vR$ for the disk stars in a given
bin, as estimated using the method described in the previous paragraph.
We use the observed $\vphi$ distribution to calculate the likelihood
distributions for ${\rm v_c}$ and ${\rm R_d}$. Given these likelihood
distributions, we can then take the moments of equation
(\ref{eq:cu94}) to estimate the mean and dispersion of $\vphi$ (and
their uncertainties).
Note that we do not infer any physical meaning from our determination of 
${\rm v_c}$ and ${\rm R_d}$ -- we are simply using the functional
form given in equation (\ref{eq:cu94}) as a convenient description.
We find that in most cases this provides a reasonable fit to the $\vphi$
distribution and should provide a more robust estimate than a sample
variance. As before, we include a non-rotating Gaussian halo component
in the fit, with the dispersion fixed at $82.4\:\kms$ \citep{Sm09}. An
example fit is shown in Figure 1.

\begin{figure}
\plotme{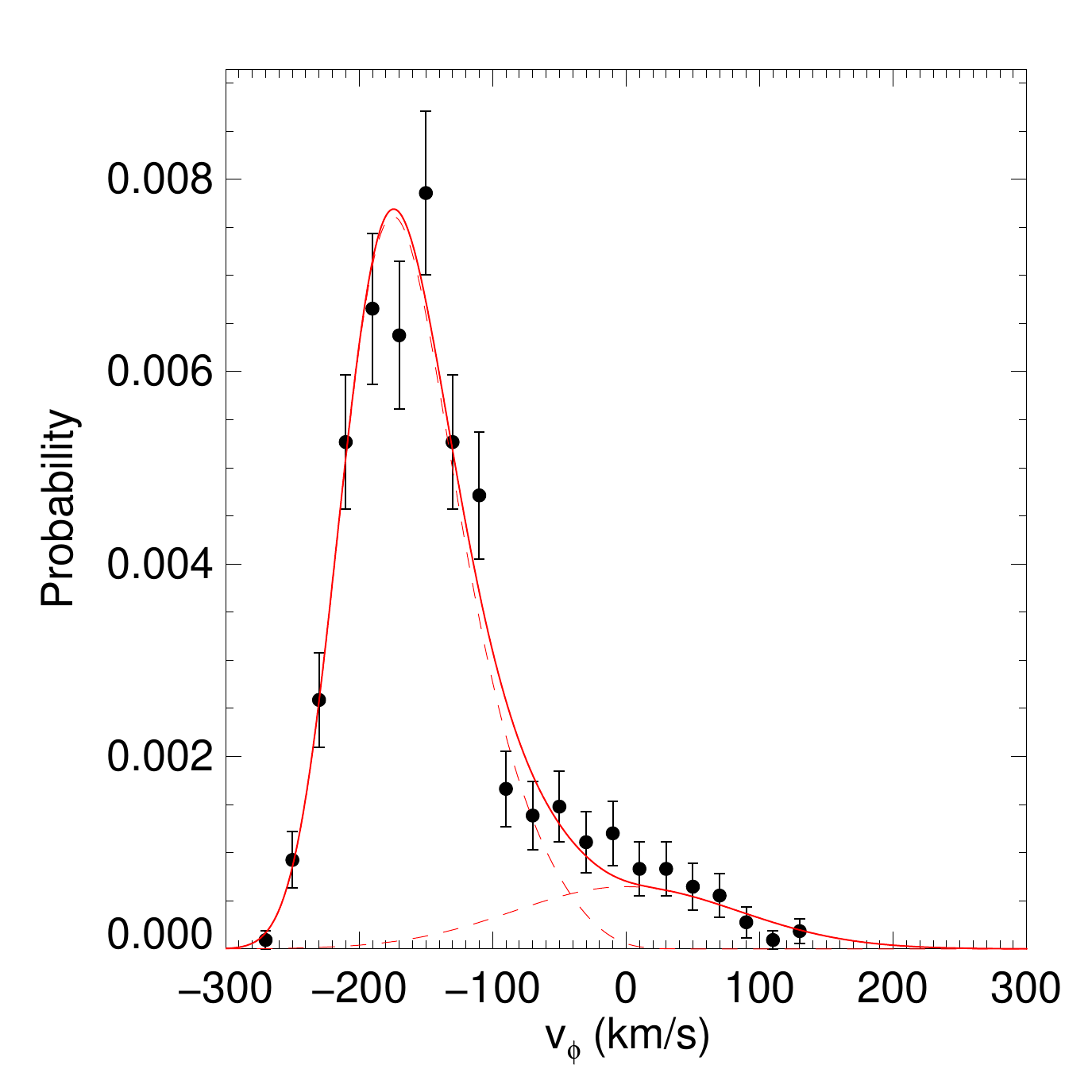}
\caption{An example of one of the fits to our distribution of $\vphi$,
  using the method described in Section \ref{sec:method_disp}. This
  figure shows the most metal-poor bin for stars at heights of around
  1 kpc below the plane (corresponding to the 10th column of Table
  \ref{tab:kinematics}).}
\label{fig:vphi_fit}
\end{figure}

\subsection{Covariance}
\label{sec:method_cov}

The covariance of the disk is calculated for each bin from the
observed covariance using the relation
\begin{equation}
\label{eq:vrvz}
\sigma_{\rm ij} = 
\langle\vi\vj\rangle_{\rm d} = \frac{\langle\vi\vj\rangle_{\rm dh} -
  f_{\rm h}\langle\vi\vj\rangle_{\rm h}}{1-f_{\rm h}},
\end{equation}
where $f_{\rm h}$ is the calculated value of the halo fraction, the
subscripts `dh' and `h' refer to the combined disk plus halo sample
and the halo respectively, and the subscripts $i$ and $j$ denote the
coordinates $R$, $\phi$ or $z$. In practice, we only analyze two
cases, namely the tilt in the ellipsoid (i.e. $\langle\vR\vz\rangle$)
and the vertex deviation (i.e. $\langle\vR\vphi\rangle$).  Note that
when we calculate $\langle\vi\vj\rangle_{\rm dh}$ we subtract the
mean, so this should explicitly be written $\langle(\vi -
\langle\vi\rangle) (\vj - \langle\vj\rangle)\rangle_{\rm dh}$.

For the halo stars, we assume that the ellipsoid is aligned in
spherical polar coordinates, or in other words, it is pointing towards
the Galactic centre. This has been shown to be a good approximation
by \citet{Sm09b}, who measured the offset to be no more than a few
degrees for a sample of halo subdwarfs.

Unlike the previous quantities, which were calculated using the
maximum likelihood method, it is not possible to incorporate the
errors on each individual velocity into this covariance
calculation. To account for the effect of the errors, we instead adopt
the following correction. For a particular star, the measured
velocity, $\vi^{\rm m}$, is equal to its true value, $\vi^{\rm t}$,
plus some displacement, $\Delta \vi$, which is due to the error in the
measurement, i.e.  $\vi^{\rm m}=\vi^{\rm t}+\Delta \vi$. Given this,
the true value of $\vi\vj$ for this star is related to the measured
value by 
\beq
\label{eq:cov_correction}
(\vi\vj)^{\rm t}=(\vi\vj)^{\rm m}-\vi^{\rm m}\Delta\vj-
\vj^{\rm m}\Delta\vi-\Delta\vi\Delta\vj.
\eeq
If we obtain the covariance by averaging this measurement over all
stars in the bin, then clearly $\langle\vi^{\rm m}\Delta \vj\rangle$
and $\langle\vj^{\rm m}\Delta \vi\rangle$ should vanish. Therefore, we
need only to correct $\langle\vi\vj\rangle$ by subtracting the term
$\langle\Delta\vi\Delta\vj\rangle$, which we can approximate using our
Monte Carlo estimates for the uncertainties (see Section
\ref{sec:sample}).

This correction is important at high $z$, where the uncertainties on
the tangential velocity are much higher. We find that this is
especially problematic for the analysis of $\covrz$, since there are a
larger fraction of stars at large $R$ (and hence with large positive
values of $\langle\Delta\vR\Delta\vz\rangle$). In order to make our
measurement more robust for $\covrz$, we reduce the radial range and
exclude any stars outside of 7.5 kpc $<$ R $<$ 8.5 kpc. This reduces
the number of stars in each bin (especially for large $z$, where some
bins now have around 300 stars) and subsequently increases our
statistical uncertainties, but we believe this is an acceptable
trade-off in order to reduce systematic errors.

Once we have calculated the covariances, we use the following formula
to obtain the angles
\begin{equation}
\label{eq:tilt}
\theta_{\rm ij}=0.5 \tan^{-1}{\left(\frac{2\sigma_{\rm ij}}{{\sigma_{\rm i}}^2 - {\sigma_{\rm j}}^2}\right)}.
\end{equation}
For both the covariances and the corresponding angles, we determine
the uncertainties using a bootstrap technique. We take 10,000
resamples from the observed distribution of $\vi$ and $\vj$, with
repetition, and calculate $\langle\vi\vj\rangle$ for each resample,
taking the dispersion of the resulting distribution as an estimate of
the uncertainty in $\langle\vi\vj\rangle$. The procedure is
similar for $\theta_{\rm ij}$, although in this case we also
incorporate the uncertainties in the dispersion, as calculated
according to Section \ref{sec:method_disp}.

\section{Kinematics of the Disk}
\label{sec:properties}

All of the analysis in this section refers to properties of the disk
population, recovered by subtraction of the halo contaminants. The
values of the kinematic properties are summarized in Table
\ref{tab:kinematics}.

\subsection{Velocity Dispersions}
\label{sec:disp}

\begin{figure}
\plotme{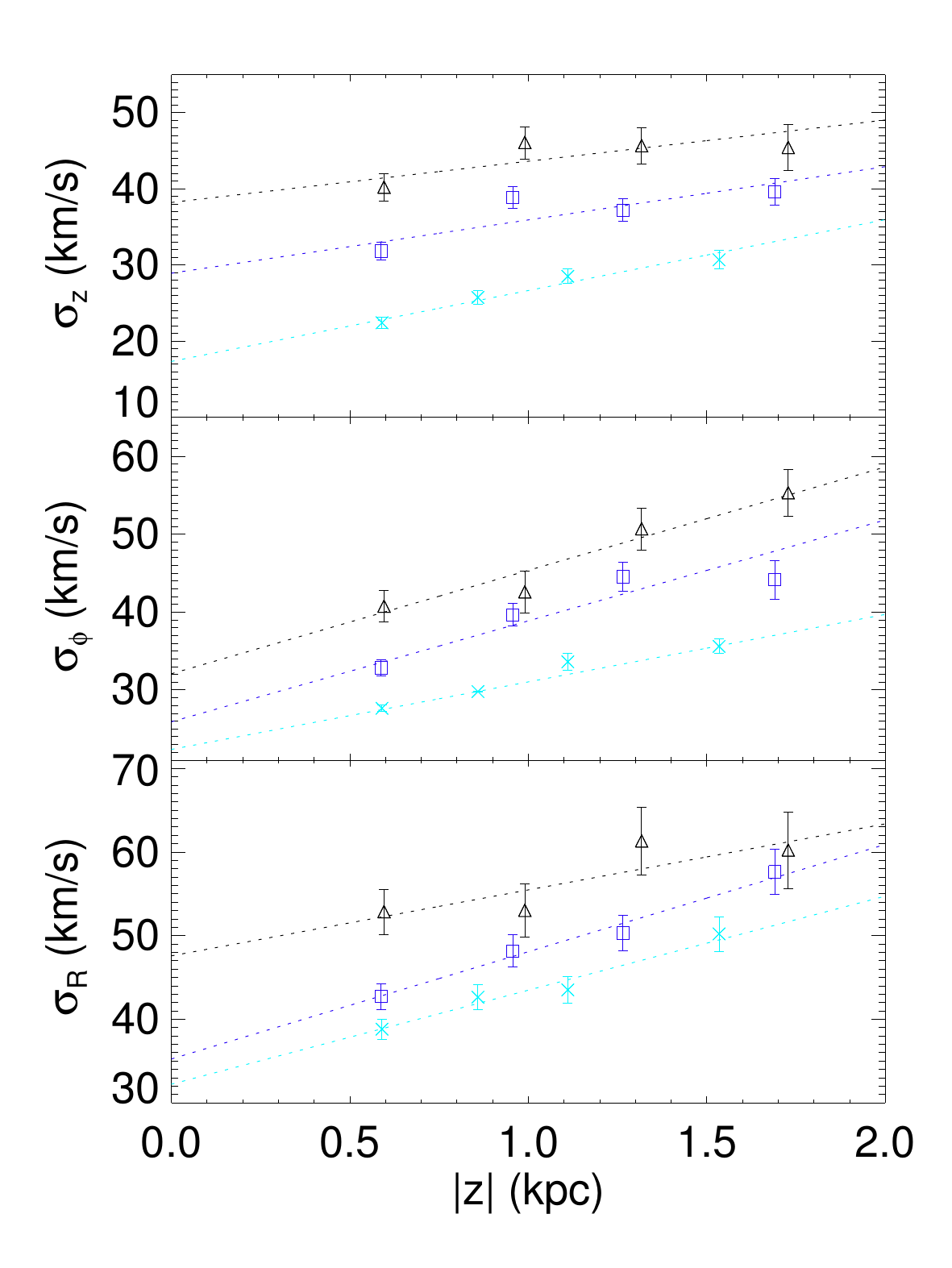}
\caption{Dispersions $\dispvr$, $\dispvz$ and $\dispvphi$ as
  functions of z and metallicity. The triangles, squares and crosses
  correspond to metallicity ranges $-1.5 \le \feh \le -0.8$, $-0.8 \le
  \feh \le -0.5$ and $-0.5 \le \feh \le 0.5$, respectively.}
\label{fig:dispersion}
\end{figure}

We plot the velocity dispersions as a function of $z$ and metallicity
in Fig. \ref{fig:dispersion}. The error bars represent the extent of
the 1-$\sigma$ confidence interval determined by the maximum
likelihood fitting. We find that populations of lower metallicity are
hotter than their more metal-rich counterparts. This is a natural
consequence of the correlation between age and metallicity - the
metal-poor stars are typically older and hence have had more time to
be affected by the heating mechanisms which cause disk stars to evolve
away from the cold, circular orbits on which they were born. We also
see that the velocity dispersions increase on moving away from the
Galactic plane, though there are hints that the velocity dispersion
$\dispvz$ may be saturating at high $z$. Table \ref{tab:disp} reports
the best-fit linear relations for these dispersions. Most are
reasonably well approximated by a linear fit, with the arguable
exception of $\dispvz$ which appears to saturate at large
$\left| z \right|$.

\begin{table}
\caption{Best-fit linear relations for the velocity dispersions.}
\begin{tabular}{lccc}
\hline
Component & $\feh$ & $\sigma \left( {\rm z}=0 \right)$ &
${\partial\sigma}/{\partial {\rm z}}$\\
& (dex) & $(\kms)$ & $({\rm \,km\,s^{-1}\,kpc^{-1}})$\\
\hline
$\dispvr$ & $\left(-0.5,+0.2\right)$ & 32.2 & 11.3 \\
$\dispvr$ & $\left(-0.8,-0.5\right)$ & 35.3 & 12.8 \\
$\dispvr$ & $\left(-1.5,-0.8\right)$ & 47.6 & 7.9 \\\\

$\dispvphi$ & $\left(-0.5,+0.2\right)$ & 22.4 & 8.7 \\
$\dispvphi$ & $\left(-0.8,-0.5\right)$ & 25.9 & 13.0 \\
$\dispvphi$ & $\left(-1.5,-0.8\right)$ & 32.1 & 13.3 \\\\

$\dispvz$ & $\left(-0.5,+0.2\right)$ & 17.4 & 9.3 \\
$\dispvz$ & $\left(-0.8,-0.5\right)$ & 28.9 & 7.0 \\
$\dispvz$ & $\left(-1.5,-0.8\right)$ & 38.2 & 5.4 \\
\hline
\end{tabular}
\label{tab:disp}
\end{table}

Such features have been found in a number of previous studies. For
example, \citet[][section 10.4.1]{BM} use a sample of nearby stars
\citep{St87} to show how the velocity dispersions vary as a function 
of metallicity. These values are consistent with what we might expect
if we extrapolate our results back to ${\rm z} = 0$. More recent
analyses have been done in the immediate solar neighborhood using
data from the Hipparcos satellite \citep[e.g.][]{DB} or the
Geneva-Copenhagen survey \citep[e.g.][]{Nordstrom04}, all of which are
consistent with an extrapolation of our results. Analyses out of the
plane have been limited due to the difficulty of obtaining reliable
six-dimensional phase-space information for distant
stars, but there are a number of such studies. For example,
\citet{Sp10} investigate trends in $\dispvphi$ with both $\feh$ and
$z$ and \citet{So08} attempt to address the relation between age and
velocity dispersion for clump giants towards the North Galactic 
Pole.
The variation in $\dispvz$ with z, which is important as it
can be used to trace the vertical potential of the disk (see Section
\ref{sec:potential}), has received a significant amount of attention,
most notably in \citet{Ku89} and more recently in \citet{Bo10}. 
\citet{Bi10} has produced a distribution function model which is able
to predict the variation in $\dispvz$ and we find that our results are
in good agreement with this, although a detailed comparison is
difficult due to the fact that we probe three separate metallicity
ranges and do not provide an overall profile.

Can we use the information contained in Fig. \ref{fig:dispersion} to
address the nature of the heating mechanisms? There are a variety of
different phenomena that could act to heat the disk, from secular
processes such as scattering due to spiral arms or molecular clouds
(see section 8.4 of \citealt*{BT} and references therein),
or external processes such as accretion of satellites onto the Milky
Way \citep[e.g.][]{Vi08}. It has been postulated that the eccentricity
structure of the thick disk can be used to discriminate formation
models \citep{Sa09,DiM10}. Recently a number of papers have
investigated this using observational data \citep{Di10,Wi10}. 
One of these works \citep{Di10} looks at the eccentricity
distribution as a function of both height and metallicity and finds
clear gradients, with the lower metallicity populations and those
further from the plane are on more eccentric orbits; this matches the
picture given by our data if we take $\dispvr$ as a proxy for
eccentricity.

\begin{figure}
\plotme{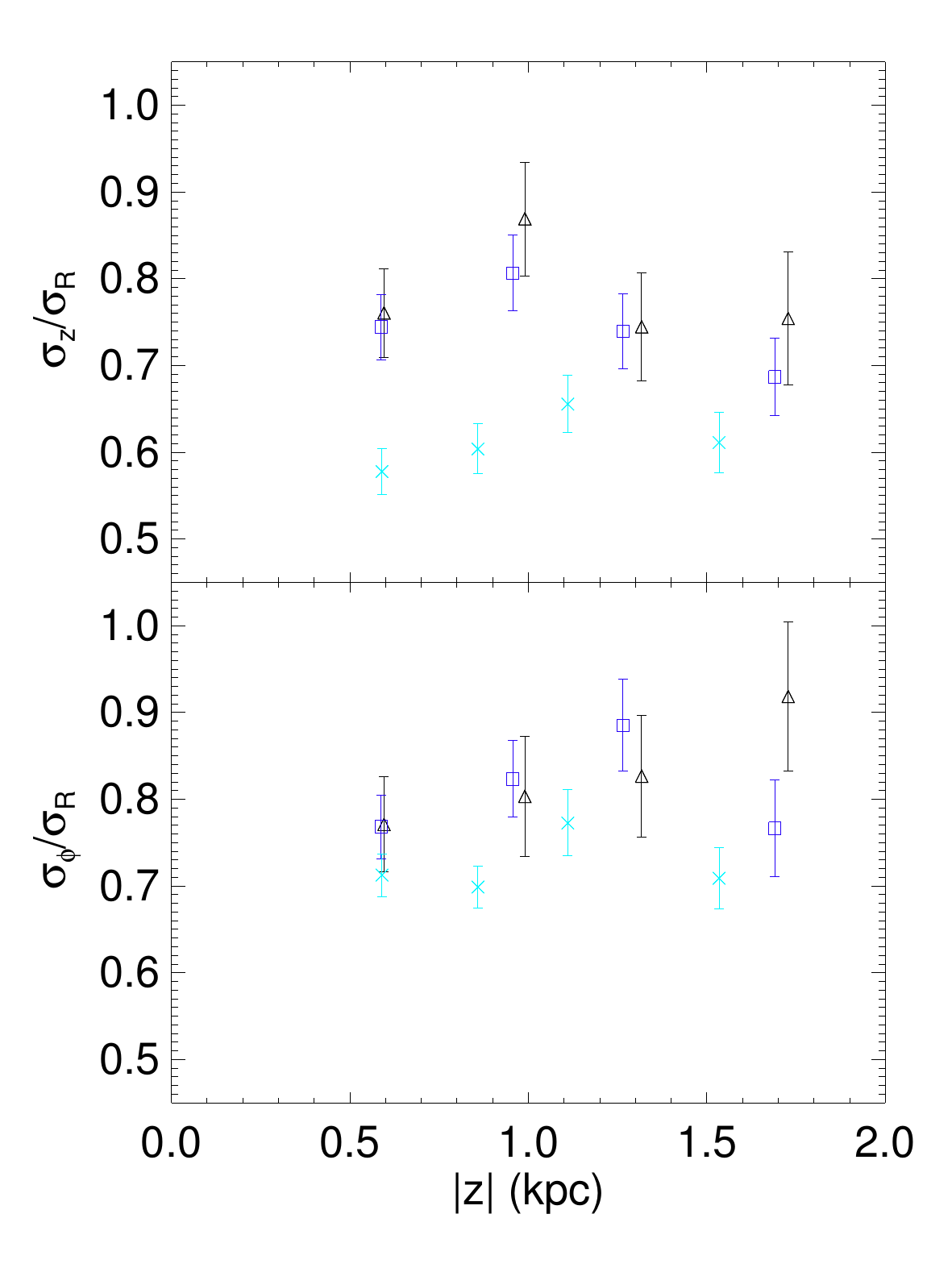}
\caption{The ratio of $\dispvz$ to $\dispvr$ and $\dispvphi$ to
$\dispvr$ as a function of z.
The triangles, squares and crosses correspond to metallicity ranges $-1.5
\le \feh \le -0.8$, $-0.8 \le \feh \le -0.5$ and $-0.5 \le \feh \le
0.5$, respectively.}
\label{fig:ratio}
\end{figure}

These heating mechanisms may also result in differences in
the predicted anisotropies for the velocity ellipsoid. In order to
investigate this, we plot the ratios of the dispersions in
Fig. \ref{fig:ratio}. The solar neighborhood estimate for
$\dispvz/\dispvr \approx 0.5$ \citep{Wi77,DB,Nordstrom04}, is
consistent with our metal-rich sample. We detect a general increase in
this ratio for lower metallicities, seemingly due to the fact that
$\dispvz$ has a stronger dependence on metallicity than $\dispvr$,
i.e. $\dispvz$ increases faster than $\dispvr$ as metallicity
decreases, resulting in larger values for $\dispvz/\dispvr$. There is
very little evidence for a gradient in $z$, and only a mild hint that
the ratio is increasing with $z$.

Theoretical studies of the expectations for $\dispvz/\dispvr$ predict
a range of values less than $1$. Classical work by \citet{Je92} looked
at heating by molecular clouds and spirals, and predicts ratios
varying from $0.4$ to $0.8$ depending on the strength of the spirals,
while a study into the effects of heating by molecular clouds alone
predicts a value of around $0.6$ \citep{Ida93}. Our results are in
keeping with these predictions.
In general, it is thought that scattering off spiral arms is
more efficient at heating in the plane, while molecular clouds are
more efficient at heating perpendicular to the plane. In this picture
one might interpret the gradient with metallicity to claim that more
metal-poor stars (i.e. older and hotter populations) are less affected
by heating from spiral arms. Whatever the reason, it is clear that the
metal-poor stars are significantly more isotropic in their kinematics
than their metal-rich counterparts.

Simulations of satellite accretion also predict similar ratios for
this ratio. For example, \citet{Vi08} and \citet{Vi10} simulate the
formation of the thick disk through heating via accretion, and also
predict a wide range of values for $\dispvz/\dispvr$ (from $\sim$ 0.4
to 0.9) depending on the parameters chosen to set-up their simulations
(e.g. satellite mass ratio, orbital inclination, etc). Therefore, it
seems that theoretical expectations cover a range of values, meaning
that our data does not simply exclude a particular mechanism. However,
now that the velocity dispersions can be found as a function of height
and metallicity, it should help to constrain future theoretical models
by excluding specific ranges of parameter space. Very few models are
able to predict the detailed behavior of these ratios and how they
vary as a function of metallicity or $z$, which will clearly become more
important as models are refined.

In recent years, there have been a number of attempts to measure
$\dispvz/\dispvr$ for external galaxies. Even though such
determinations are model dependent and based on a number of
assumptions, it is interesting to compare these to the Milky
Way. \citet{Ge97,Ge00} looked at the large spiral galaxies NGC 488 and
NGC 2985, finding ratios of $0.7$ and $0.8$, respectively,
i.e. slightly larger than the classical Milky Way determination of
$0.5$. Similar results were also found by \citet{vdK99} in their study
of a compilation of around 40 edge-on disk galaxies, with typical
values ranging from $0.5$ to $0.7$ and with no evidence for a
correlation between these values and the galaxy size or morphological
classification (from Sb to Sd).  Although larger than the Milky Way
value, these ratios for external galaxies are consistent with our
data; as can be seen in Fig. \ref{fig:ratio}, as one departs from the
plane and moves to lower metallicities values of $\dispvz/\dispvr =
0.5-0.8$ are typical.

The ratio $\dispvphi/\dispvr$ also provides information about heating
processes, though this has received less attention in the
literature. The value in the solar neighborhood is of interest as it
is related to the slope of the rotation curve of the Galaxy. The ratio
$\dispvphi^2/\dispvr^2$, which is sometimes refers to as Oort's ratio,
is predicted to lie around 0.5 to 0.6 depending on the shape of the
rotation curve (\citealt*{KuT91}; see also \citealt{EC93}). An
extrapolation back to the solar neighborhood gives a good agreement
with this expectation, similar to the results of \citet{DB}. In
Fig. \ref{fig:ratio}, we can see that the gradient with metallicity is
less prominent than for $\dispvz/\dispvr$, although it also looks like
in general the metal-poor stars are more isotropic.  As with
$\dispvz/\dispvr$, there is only weak indication that stars further
from the plane exhibit larger ratios; in particular it is worth noting
that the distribution for the metal-rich stars is essentially flat.

%\remark{Is there anything interesting we can do with
%$\dispvphi/\dispvr$ - This term is discussed pretty extensively in
%\citet{Cu94}.}

%%%%%%%%%%%%%%%%%%%%%%%%%%%%%%%%%%%%%%%%%%%%%%%%%%%%%%%%%%%%%%%%%%%%%%%%%


%%%%%%%%%%%%%%%%%%%%%%%%%%%%%%%%%%%%%%%%%%%%%%%%%%%%%%%%%%%%%%%%%%%%%%%%%

\subsection{The Rotation Lag}
\label{sec:ASdrift}

It has long been understood that there is a correlation between the speed
at which a population of stars rotates around the Galactic centre and
the velocity dispersion of this population. This is known as the
asymmetric drift \citep[e.g. section 10.3.1 of][]{BM} and is clearly
evident in the solar neighborhood \citep[e.g.][]{DB}. The behavior
out of the plane is more controversial. Whilst the existence of a
correlation between lag and height from the plane has been observed in
numerous studies \citep[e.g.][]{Wy86,Ma92,Ch00,Gi06}, the situation
regarding trends with metallicity are less clear. In recent years
there have been at least two papers with seemingly irreconcilable
views; \citet{Sp10} claim to have detected a large gradient, at a
level of around 40 to 50 $\kms$ dex$^{-1}$, while \citet{Iv08} claim
an essentially flat relation with no detectable gradient.

\begin{figure}
\plotme{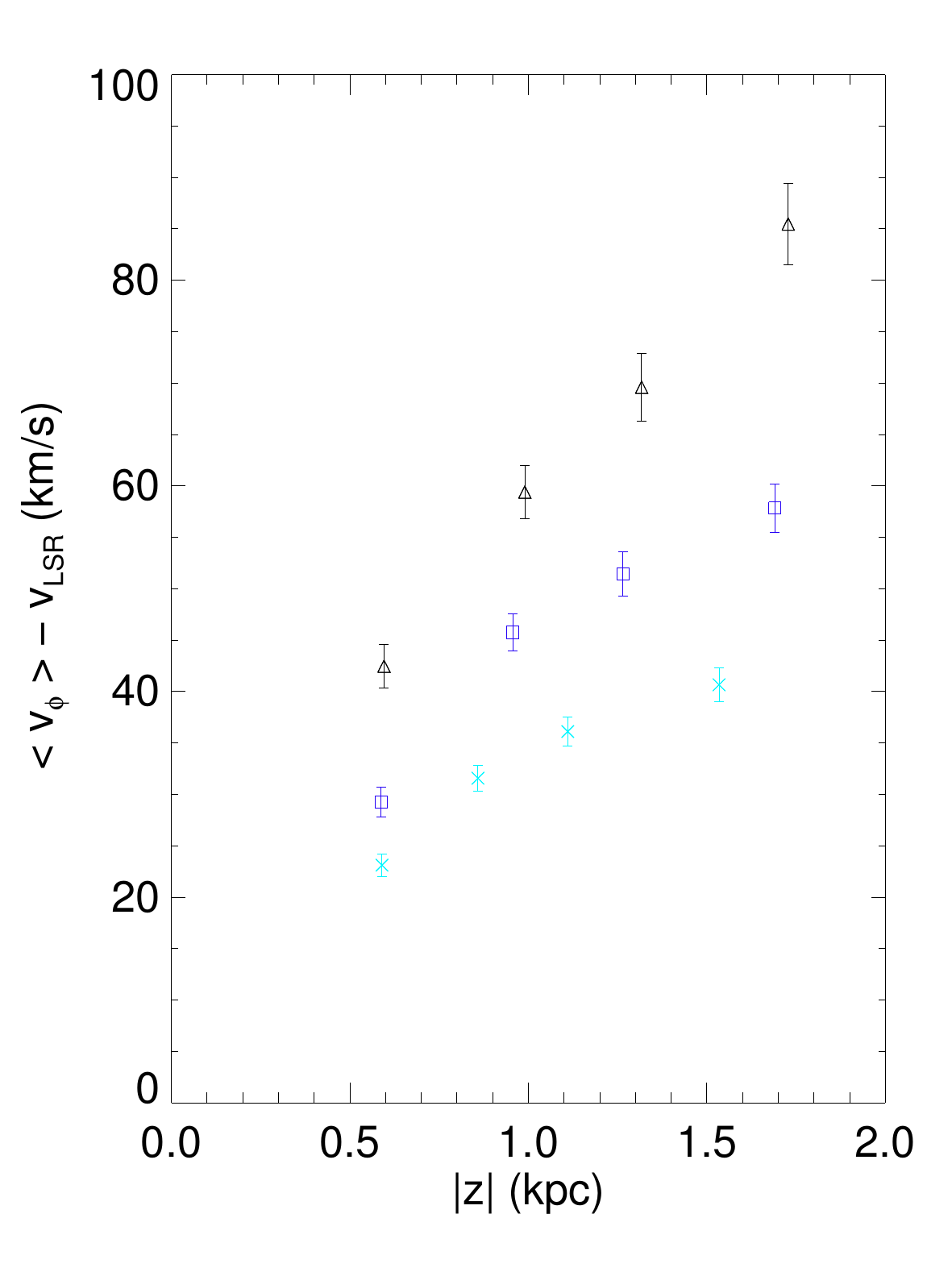}
\caption{The rotational lag, plotted against z. Velocities are
  relative to the Local Standard of Rest, assumed to be as determined
  by \citet{DB}, which means that in this plot the Sun would lie at
  (0, -5 $\kms$). The triangles, squares and crosses correspond to
  metallicity ranges $-1.5 \le \feh \le -0.8$, $-0.8 \le \feh \le
  -0.5$ and $-0.5 \le \feh \le 0.5$, respectively.}
\label{fig:lag_z}
\end{figure}

Our findings are presented in Fig. \ref{fig:lag_z}, where it is
immediately evident that there are clear trends in the lag. The hotter
populations (i.e. those with lower metallicity or those further from
the plane) exhibit significantly more lag than their colder
counterparts, with the metal-poor stars rotating more
than 80 $\kms$ slower than the LSR at 2 kpc from the plane.
The gradient of the lag with respect to $|$z$|$ varies
from around 15 to 40 $\kms{\rm kpc}^{-1}$, depending on
metallicity. This is in agreement with many of the literature values
which typically find values around 30 $\kms{\rm kpc}^{-1}$
\citep[e.g.][]{Gi06}. It is curious to note that for the more
metal-rich stars there appears to be a saturation in the level of the
lag as one moves from the plane; from Fig. \ref{fig:lag_z} it can be
seen that while the metal-poor stars exhibit an almost linear trend,
the gradient for the metal-rich and intermediate-metallicity
populations becomes shallower for z $\ga 1$ kpc.

Despite the strong theoretical grounds for expecting the lag to be
larger for hotter (and hence more metal-poor) populations, as is seen
in our data, some authors disagree with this claim. In particular,
\citet{Iv08} have argued that such a trend with metallicity is not
observed in a sample of SDSS data. Their viewpoint is seemingly lent
weight by the findings of \citet{Lo10}, who analyzed an N-body+SPH
disk galaxy simulation and found no clear trend when plotting rotation
velocity versus metallicity (see fig. 10). However, this does not imply
that such a trend is entirely absent from their simulations; their
fig. 9 shows that young stars do indeed possess larger rotation
velocities compared to the older stars and also that there is a clear
age-metallicity gradient. If instead \citet{Lo10} slice their data in
the same manner as we have (i.e. taken four bins in $|$z$|$ from $0.5$
to $2$ kpc and three bins in metallicity), then they see behavior
which is qualitatively identical to our Fig. \ref{fig:lag_z} with a
clear gradient in metallicity (V. Debattista, private communication).

\begin{figure}
\plotme{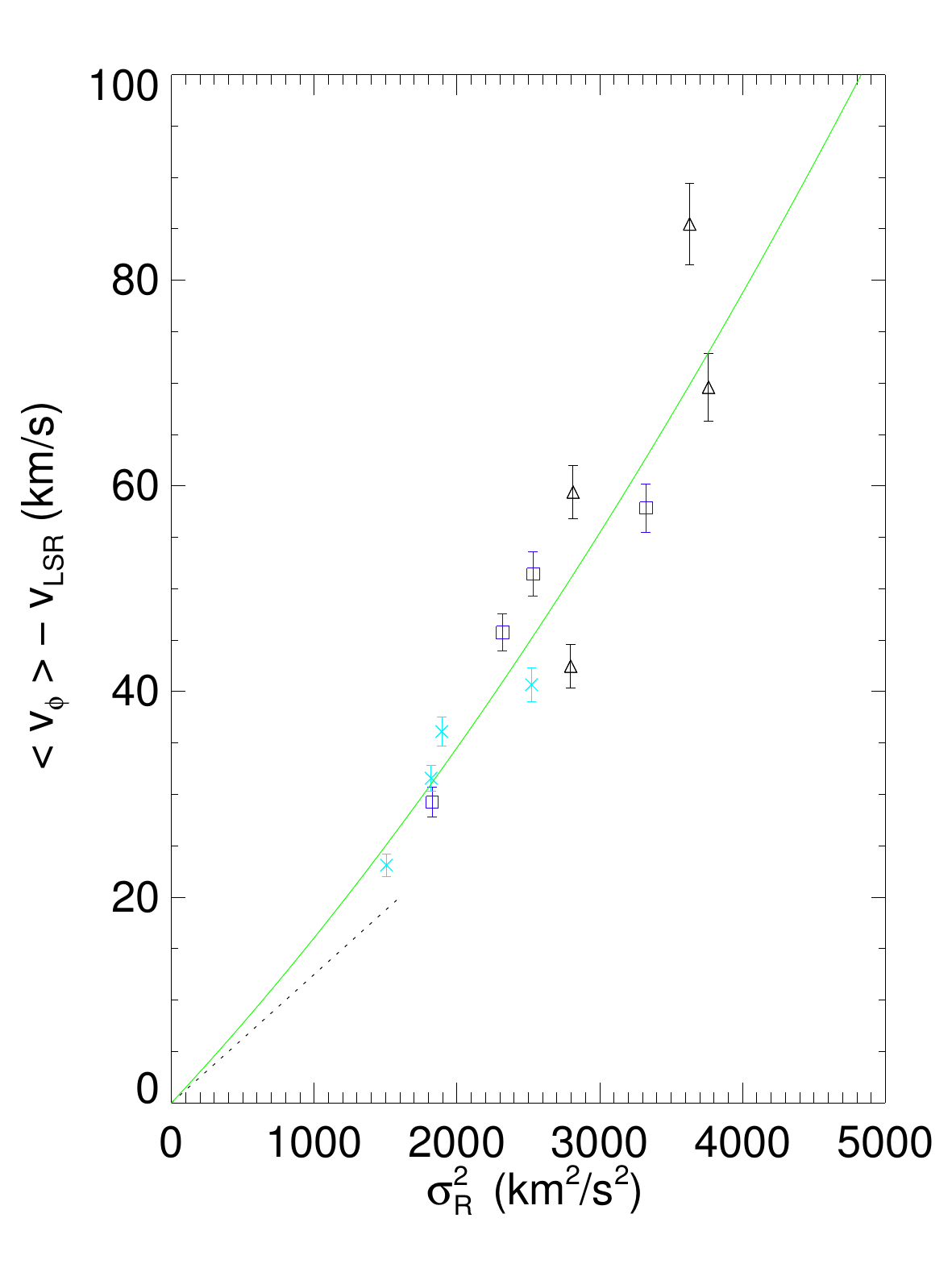}
\caption{Rotational lag as a function of radial velocity
dispersion. The triangles, squares and crosses correspond to metallicity
ranges $-1.5 \le \feh \le -0.8$, $-0.8 \le \feh \le -0.5$ and $-0.5
\le \feh \le 0.5$, respectively. The dotted line corresponds to the
solar-neighborhood relation from \citet{DB}. The solid line denotes
an empirical fit with the lag equal to $0.0149 \dispvr^2 + 1.21 \times
10 ^{-6} \dispvr^4$.}
\label{fig:lag_sig}
\end{figure}

In the solar neighborhood, \citet{DB} found that there is a linear
relation between lag and the square of the radial velocity dispersion,
which is in good agreement with theory \citep[e.g. section 4.8.2a
of][]{BT}. However, it is less clear what is expected as we depart from
the plane. We show the observational result for our data in
Fig. \ref{fig:lag_sig}. As in the solar neighborhood, there appears
to be a relatively tight correlation with $\vrsq$ that is independent
of metallicity. However, the gradient is much steeper than that of the
solar neighborhood, where it is found that $\langle\vphi\rangle-\vc =
\langle\vR^2\rangle/(80\pm5\:\kms)$. The relation is reasonably well-fit
by an quadratic relation, as shown in Fig. \ref{fig:lag_sig}, where we
find,
\beq
\langle \vphi \rangle - \vc = 0.0149 \dispvr^2 + 1.21 \times 10 ^{-6} \dispvr^4.
\eeq

Note that our analysis has been carried out using the \citet{DB} value
for the Sun's motion with respect to the local standard of rest. This
has recently been revised by \citet{Bi10}, with the velocity in the
direction of rotation being increased from 5.2 $\kms$ to 11 $\kms$. If
we adopt this newer value, then it implies that the lag which we
measure will be too large by around $\sim 5\:\kms$. However, since
this offset should be applied uniformly to all of our sample, the
gradients in lag with metallicity and height from the plane, which we
can see from Fig. \ref{fig:lag_z}, will be unaffected.

\subsection{Radial and Vertical Bulk Motions}

\begin{figure}
\plotme{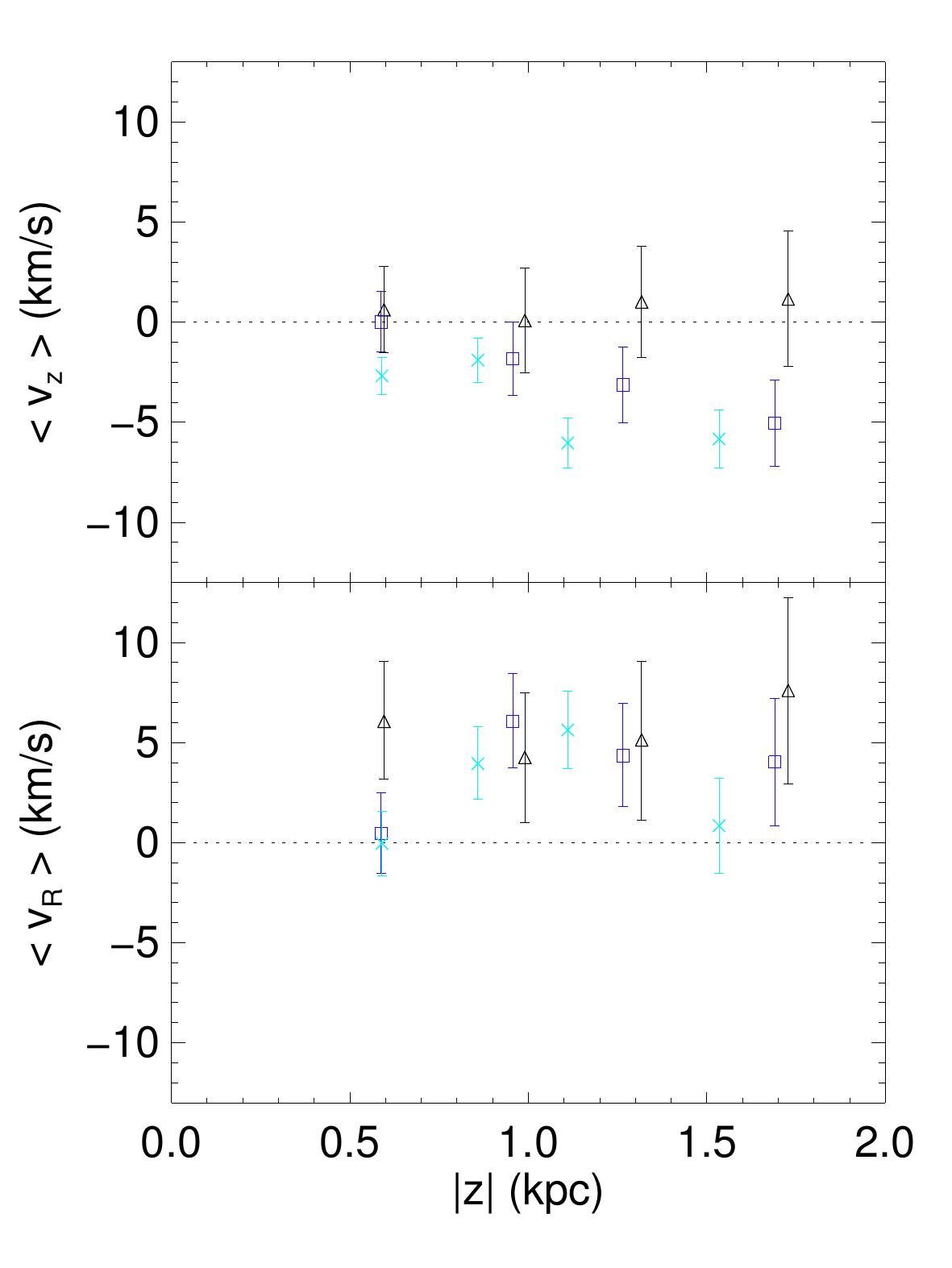}
\caption{Mean velocities as a function of height from the plane.
The triangles, squares and crosses correspond to disk
stars with metallicities $-1.5 \le \feh \le -0.8$, $-0.8 \le \feh \le
-0.5$ and $\feh \ge -0.5$, respectively.}
\label{fig:means}
\end{figure}

The bulk motion for the other two components, $\vz$ and $\vR$ are
presented in Fig. \ref{fig:means} and Table
\ref{tab:kinematics}. Curiously $\langle \vz \rangle$,
whilst appearing consistent with zero for the metal-poor stars, seems
to deviate from zero on moving away from the plane for the more
metal-rich samples. Although tentative, this is an intriguing result
and, if confirmed by further observations, could have important
implications.

In the lower panel of Fig. \ref{fig:means}, we see that the radial
component also seems to exhibit a net motion, with the sample moving
outwards from the Galactic centre. This has also been seen in data
from the RAVE survey \citep{Si10}, using a sample of stars with
$|z|<1$ kpc. Here, we seem to be detecting similar behavior out to
$|z| \sim 2$ kpc, though there are no clear trends with either $|$z$|$
or metallicity. This is arguably surprising, since \citet{Si10}
suggested that such behavior should be due to non-axisymmetric
components in the Galaxy, such as the bar or spiral arms, which one
might expect to have a greater effect on stars in the plane. The other
potential source of non-axisymmetry postulated by \citet{Si10} was
that of ellipticity in the outer dark matter halo, which we might
expect to lead to a lack of a gradient with $|$z$|$.

However, there are difficulties associated with measuring such
quantities and various systematic biases could affect these results,
for example mistaken assumptions regarding the Local Standard of Rest
or the Sun's peculiar motion.

%%%%%%%%%%%%%%%%%%%%%%%%%%%%%%%%%%%%%%%%%%%%%%

\subsection{Tilt of the Velocity Ellipsoid}

\label{sec:covariances}

\begin{figure}
\plotme{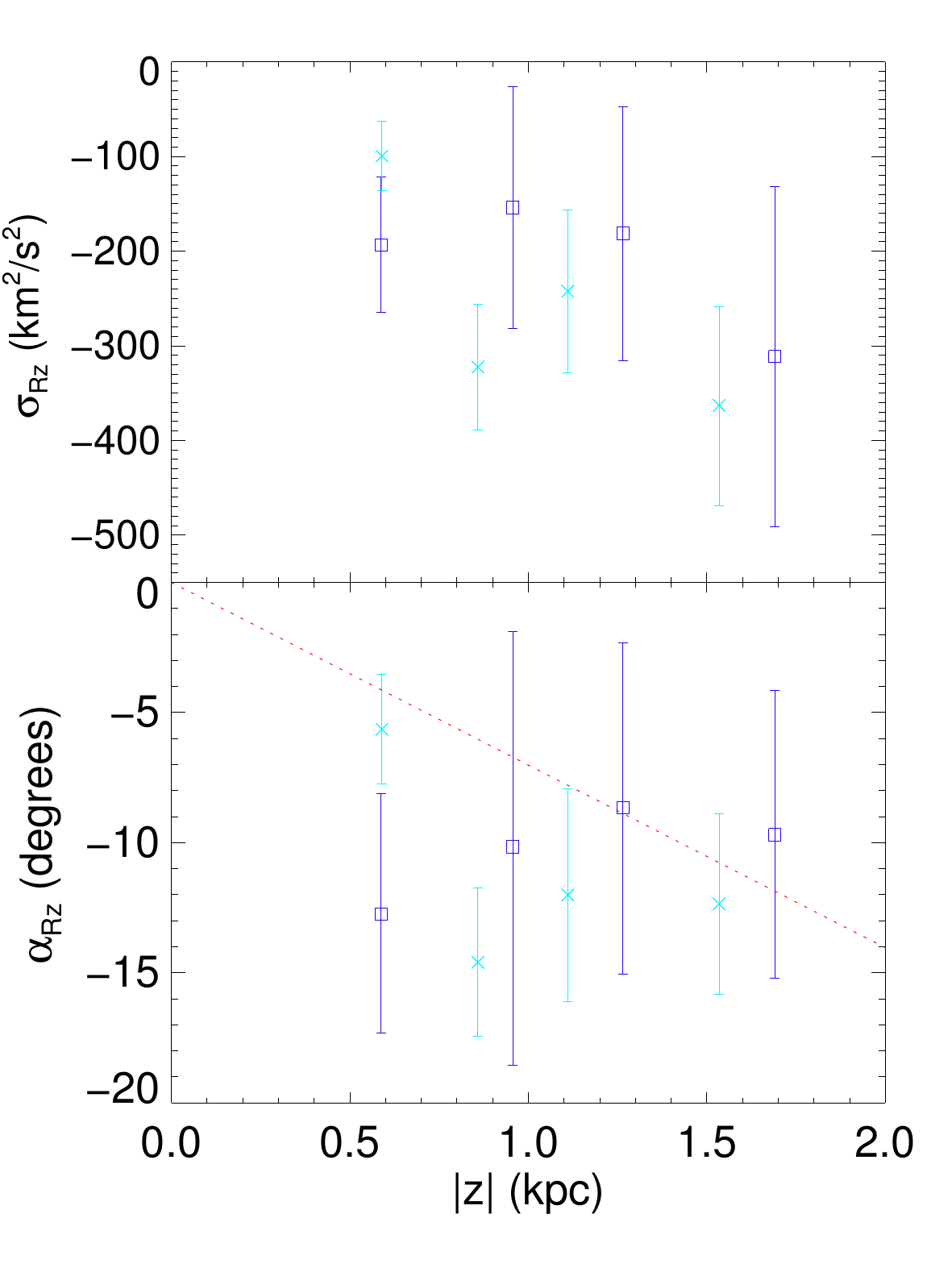}
\caption{The variation of $\covrz$ and the corresponding angle
$\alpha_{\rm Rz}$, often referred to as the tilt, as a function of
height from the plane.  The dashed line is the assumed halo tilt
(i.e. aligned in spherical polars).  The squares and crosses
correspond to disk stars with metallicities
$-0.8 \le \feh \le -0.5$ and $\feh \ge -0.5$, respectively.}
\label{fig:vrvz}
\end{figure}

The covariance between $\vphi$ and $\vz$, which is often referred to
as the tilt term, is an important quantity, which we can attempt to
determine with our data. Little is known about how this behaves when
one moves away from the Galactic plane. At $z =0$, we expect this to
vanish due to symmetry arguments, but the behavior out of the plane
is harder to measure as it requires distances to be determined to high
accuracy. Despite these difficulties, the orientation of the ellipsoid
is of importance for understanding the shape of the potential 
\citep[c.f.][]{Sm09b,Bi11,Ev11}. Theoretical
predictions span a range of values lying anywhere between velocity
ellipsoids parallel to the Galactic plane to ones which point towards
the Galactic centre \citep{Bi83,Ku89}. The consensus \citep[e.g.][]{BT}
seems to be that an alignment midway between cylindrical polar and
spherical polar is most reasonable.

One of the recent determinations of the tilt was by \citet{Siebert08}.
They used red clump giants from the RAVE survey to measure a tilt
angle of $7^\circ.3\pm1^\circ.8$ at $z\sim 1$ kpc below the plane,
with the orientation inclined towards the Galactic centre. Their
sample of stars covers a large range in $z$ (from 0.5 to 1.5 kpc),
which complicates the interpretation of their result, but it is clear
that this value is close to what one would expect for an ellipsoid
aligned with the centre of the Galaxy (i.e. ${\rm tan}^{-1} 1/8 =
7^\circ.1$). This is an unexpected result as it lies at the
extremum of the predicted values, but it appears to be robust as it has
been supported by subsequent studies \citep{Ca11}.

Our results are given in Table \ref{tab:kinematics}. We have omitted
the values for the metal-poor stars due to problems in accurately
measuring their covariances. This is because it is very difficult to
reliably correct for the halo contamination - for the metal-poor
sample the covariance measurements are dominated by the contribution
from the halo stars (since $\covrz$ for the halo is much larger than
for the disk) and so reliably extracting the disk covariance is
practically impossible. We include the medium-metallicity sample in
our analysis as the the halo contamination is smaller and hence our
determination more robust.

The covariances are plotted in Fig. \ref{fig:vrvz}. There appears to
be a weak trend, with the magnitude of $\covrz$ increasing
slightly as we move away from the plane. If we now convert the
covariances into an angles using equation (\ref{eq:tilt}), we obtain
the results shown in the lower panel of Fig. \ref{fig:vrvz}. The
dotted line corresponds to what we would expect for a velocity
ellipsoid aligned in spherical polar coordinates, and so we would
expect our data to lie between this dotted line and the
$\alpha_{\rm Rz}=0$ axis. Although there are large uncertainties, it
appears that the metal-rich and medium-metallicity 
stars are in general consistent with the dotted line (and hence
consistent with the \citealt*{Siebert08} result). A couple of points
are more than one-sigma away from this line, but we believe this is
probably an artifact; it is very difficult to explain such
behavior for a disk in equilibrium, which implies that either our
disk is not in equilibrium (due to accretion remnants affecting the
distribution, or due to transient effects from resonances associated
with spiral arms or the bar) or our data are suffering from
observational biases.

To conclude, although this is a difficult measurement with large
uncertainties, it does appear that the tilt angle for the disk is
surprisingly large, reinforcing the findings of \citet{Siebert08}.

\subsection{The Vertex Deviation}

The orientation of the velocity ellipsoid in the $(\vR,\vphi)$-plane,
often referred to as the vertex deviation, is also of interest. In the
immediate solar neighborhood, this has been found to be around
$10^\circ$ to $20^\circ$ \citep{DB}, with corresponding covariance
$\covrphi \sim 100$ km$^2$s$^{-2}$. For an axisymmetric system this
term should vanish and the fact that it is non-zero is usually
attributed to resonances from either the Galactic bar or spiral
arms \citep[e.g.][]{De99,Qu03,DeS04,Minchev10b}.

With our data we are able to probe the vertex deviation out of the
plane. Unlike the tilt angle discussed in the previous section, the
vertex deviation is somewhat easier to measure. Since our data are
located towards the South Galactic cap, velocities in the
$(\vR,\vphi)$-plane are mostly derived from the proper-motion data
and, as a consequence, correlated uncertainties are less of a
problem.
%and hence are not as strongly correlated as those in the
%$(\vR,\vz)$-plane (where the $\vR$ component comes from the proper
%motion and the $\vz$ component comes from the line-of-sight velocity,
%resulting in strong correlations).
The halo correction is also easier to handle, although it still makes
it difficult to robustly measure $\covrphi$ for the metal-poor
sample. Despite the fact that the halo $\covrphi=0$, the mean $\vphi$ is
significantly offset from that of the disk and the dispersions are
large. Therefore, unless the halo velocity distributions are well
sampled (which they are not), there are likely to be large statistical
fluctuations due to the influence of the halo contamination between
different bins, even though the mean correction will be zero.

\begin{figure}
\plotme{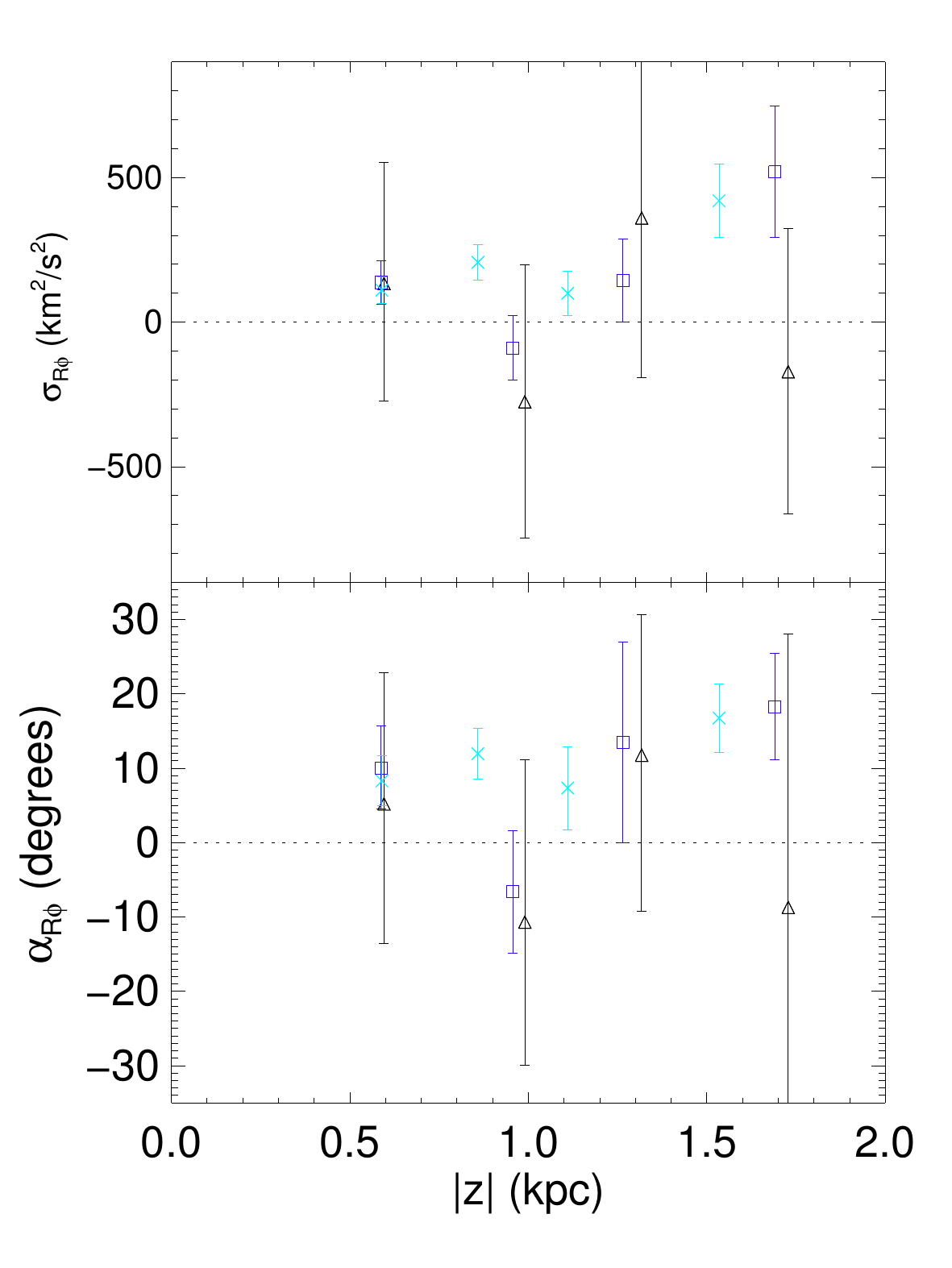}
\caption{The variation of $\vrvphi$ and the corresponding angle 
$\alpha_{\rm R\phi}$, often referred to as the vertex
deviation, as a function of height from the plane.
The triangles, squares and crosses correspond to disk
stars with metallicities $-1.5 \le \feh \le -0.8$, $-0.8 \le \feh \le
-0.5$ and $\feh \ge -0.5$, respectively.}
\label{fig:vertex}
\end{figure}

Our results are presented in Fig. \ref{fig:vertex} and Table
\ref{tab:kinematics}. As expected, there
are large fluctuations in the measurement of $\covrphi$ for the most
metal-poor stars For the rest, there is a clear positive signal, which
is consistent with the value found by \citet{DB}. If we concentrate on
the most metal-rich stars, which have much smaller errors than the
more metal-poor stars, the distribution appears to be flat (assuming
$\covrphi=100$ ${\rm km^2s^{-2}}$ at z=0) with a small increase for
bin furthest from the plane. A similar rise is also seen for the
medium-metallicity sample. Instead of this weak positive gradient, we
might naively expect that $\covrphi$ should $decrease$ with increasing
height from the plane, since this offset is thought to be due to
features which are stronger in the plane (namely the bar and spiral
arms). However, this is only a very weak detection and so further
study is required before any definitive statements can be made.

\section{An Application: The Galactic Potential}
\label{sec:potential}

There are many applications of the data we have provided. We now show
one such example where we use the vertical dispersions to compute the
Galactic potential close to the disk.

For an isothermal disk it can be shown that,
\beq
\frac{1}{\nu}\frac{\partial(\nu\dispvz^2)}{\partial{\rm z}}
=
-\frac{\partial\Phi}{\partial{\rm z}}
\label{eq:isothermal}
\eeq
To obtain this relation we have neglected the contribution of the tilt
term ($\covrz$); even if the tilt is pointing towards the Galactic
centre (as suggested by Fig. \ref{fig:vrvz}), then the contribution of
these neglected terms to equation (\ref{eq:isothermal}) will be at a
level of around five per cent \citep[see equation 4.271
of][]{BT}. From equation (\ref{eq:isothermal}) it is trivial to show
that,
\beq
\nu({\rm z}) = \nu(0) \, {\rm exp} \left[ \frac{\Phi(0)-\Phi({\rm z})}{\dispvz^2} \right].
\label{eq:iso_pot}
\eeq

However, it is clear that the disk is not isothermal. If we generalize
equation (\ref{eq:isothermal}) to a population with dispersion
$\dispvz({\rm z})$ then we obtain the following formula,
\beq
\frac{ \nu({\rm z}) } { \nu(0) } =
\frac{ \dispvz^2(0) }{ \dispvz^2({\rm z}) }
\, {\rm exp} \left[ 
\int d{\rm z}
\frac{1}{\dispvz^2}
\frac{\partial\Phi}{\partial{\rm z}}
\right].
\label{eq:pot_full}
\eeq
By taking a potential of the form,
\beq
\Delta\Phi = \Phi({\rm z}) - \Phi(0) = a \left| {\rm z} \right|
+ b \left| {\rm z} \right|^2 + ...
\label{eq:pot_expansion}
\eeq
and retaining only the first two terms, we obtain the following relation,
\beq
\nu({\rm z})\,\dispvz^2({\rm z}) = \nu(0)\,\dispvz^2(0)
+ \int_0^{\rm z} d{\rm z^\prime} \nu({\rm z}^\prime)
\left[
a + 2b{\rm z}{^\prime}
\right].
\label{eq:pot}
\eeq

\begin{figure}
\plotme{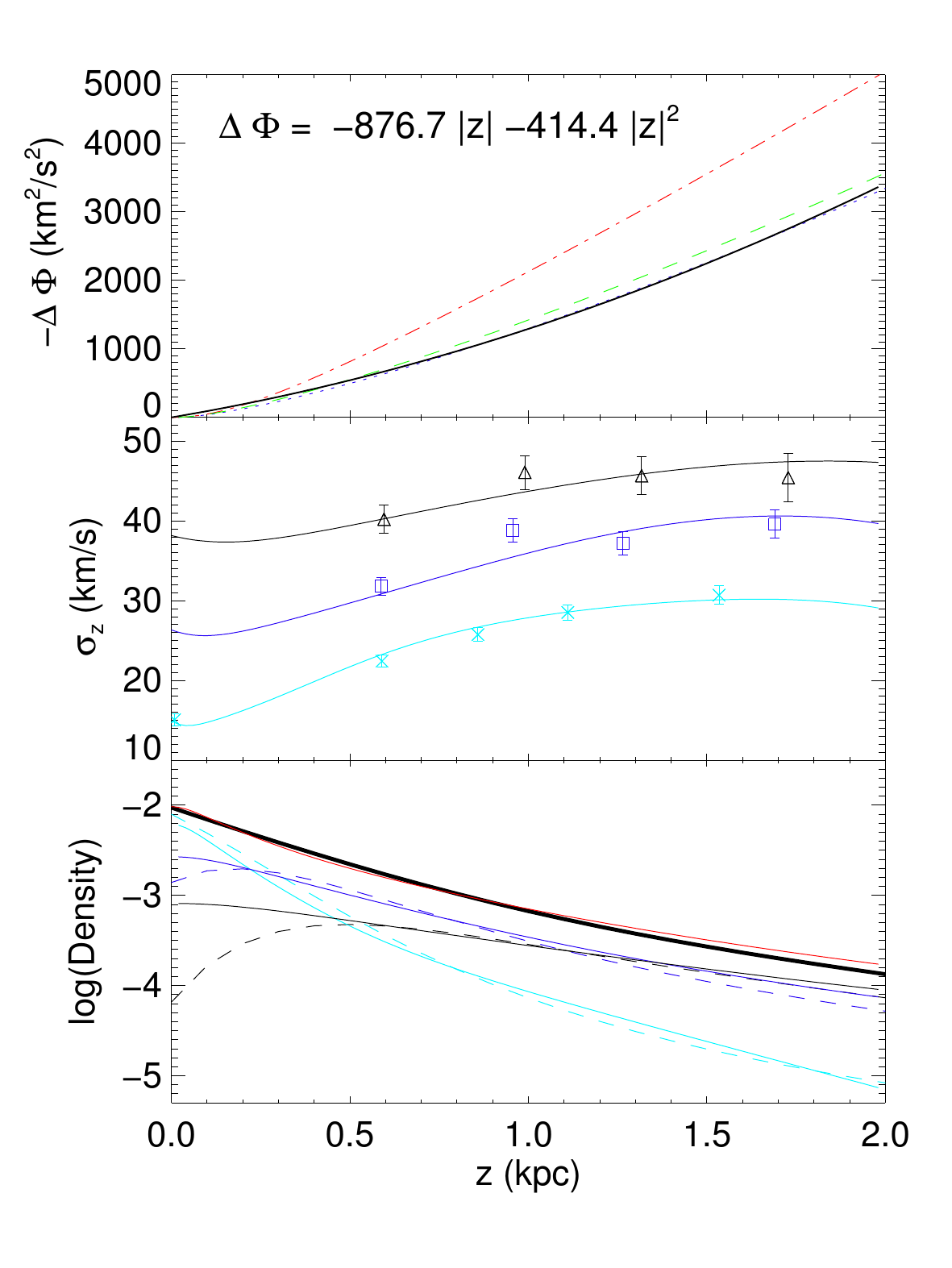}
\caption{The results of the application to constrain the potential of
  the Galactic disk, as described in Section \ref{sec:potential}. The
  bottom panel shows the density distribution. In the range 
  $0.5 < {\rm z/kpc} < 2$ we fit sech profiles to the metal-rich
  (cyan), medium-metallicity (blue) and metal-poor (black) profiles
  derived from the work of Juri\'c/Ivezi\'c. In the range ${\rm z/kpc} <
  0.5$ we fit to the total density profile (thick black line). The
  total of our three density components is given by the thin red
  line. The middle panel shows the fit to the dispersion profiles,
  with the data taken from Fig. \ref{fig:dispersion} plus an additional
  metal-rich point at ${\rm z} \approx 0$ from the Geneva-Copenhagen
  survey. The upper panel shows the potential resulting from these
  fits. For the purposes of comparison, included in the upper panel
  are models for the potential taken from literature sources, namely
  \citeauthor{DB98b} (1998 -- Model 1, dotted; Model 4, dashed) and
  \citeauthor{Fe06} (2006 -- dot-dashed).}
\label{fig:potential}
\end{figure}

This equation allows us to constrain the potential of the disk using
our measured dispersions, provided we know the form of the density
distribution for our tracer populations, i.e. our three metallicity
ranges. Unfortunately this proviso is not an easy one to overcome. The
complicated nature of the SDSS spectroscopic selection function
prohibits us from using our current spectroscopic dataset to determine
the density distribution. Therefore we resort to photometric number
counts, using the overall density distribution of \citet{Juric08},
convolved with the metallicity distribution function of \citet[][as
revised in the appendix of \citealt*{Bo10}]{Iv08}. We cannot use these
density distributions directly as they provide unrealistic predictions
for the behavior at small z where, for the intermediate- and
low-metallicity ranges, the resulting density actually rises as z
increases (see the lower panel of Fig. \ref{fig:potential}). To remedy
this we assume that each of our three components follow a sech to the
power 0.4 profile \citep{Ba07} and fit these functions to the Juri\'c/Ivezi\'c
profile over the range $0.5 < {\rm z/kpc} < 2$. For the range $0 <
{\rm z/kpc} < 0.5$, where the individual profiles are unreliable, we
require the sum of our three density distributions to match the
Juri\'c total density distribution. We therefore include the
scale-heights and normalizations of the three components as free
parameters in our fit. Since the Juri\'c disk profile consists of a
thin- and thick-component, we similarly assume each of our metallicity
populations consists of two components, keeping the scale-height 
ratio and the normalization of the thin- and thick-component fixed at
the values adopted by Juri\'c (${\rm z_{h,thick}/z_{h,thin}} = 3$ and
$\nu_{\rm thick}(0)/\nu_{\rm thin}(0) = 0.13$). 

This means that we have a total of 11 free parameters: two to describe
the potential plus three parameters per metallicity range
($\dispvz(0)$ and a normalization and scale-height for the
thin-component of the density profile). We then fit simultaneously the
Juri\'c/Ivezi\'c density profiles and the dispersion profiles (via
equation \ref{eq:pot}) by using a standard $\chi^2$ technique. We
assume that the errors on the density profiles scale with $\sqrt{\nu}$.

Since our data do not constrain the dispersion profile for ${\rm z}
\la 0.5$ kpc, we add an additional metal-rich datapoint for the
immediate solar-neighborhood using data from the Geneva-Copenhagen
survey \citep{Nordstrom04,Ho09}. We take the 252 stars with $\feh >
-0.5$ dex, parallax error less than 13 per cent and distances less
than 100 pc and find that the $\dispvz = 15.1 \pm 0.7\:\kms$.

As we are simultaneously fitting both the density and the dispersion
profiles, it is in some sense arbitrary how we weight these two
components of the fit. We have chosen to construct our fit so that the
overall $\chi^2$ is approximately equal for each of the two components,
which implies that our fit is not dominated by either the density or
dispersion profiles.

The resulting best fit is shown in Fig. \ref{fig:potential} with parameters
given in Table \ref{tab:potential}. Although the
Juri\'c/Ivesi\'c relations are not necessarily well represented by our
sech profiles, note that the total density distribution from our three
components provides a good match to the overall \citet{Juric08}
distribution. However, it should be noted that the fit relies strongly
on our assumed Juri\'c/Ivezi\'c density profiles, which are subject to
uncertainties that may bias our result. 
For example, one clear failing of our model is that the
solar-neighborhood normalizations do not match observations. In
particular $\nu(0)$ for the metal-rich sample is only around twice that
of the intermediate-metallicity sample, whereas we know that in the
solar-neighborhood stars with $\feh > -0.5$ make up around 95 per
cent \citep{Ho09}. As a consequence our resulting potential should not
be over-interpreted. 

Given these caveats, it is reassuring to see that our potential (shown
in the top panel of Fig. \ref{fig:potential}) is in good agreement
with existing models. Most notably our simple model is in
exceptional agreement with the models of \citet[][hereafter DB]{DB98b},
especially their Model 1. Although there is a slight
discrepancy\footnote{This discrepancy is most-likely due to the fact
that in our simple model for the potential (equation
\ref{eq:pot_expansion}) one component of the mass distribution is an
unphysical infinite razor-thin sheet. A better match at small z could
be found if one distributed this mass in a more realistic manner.} at
small z (${\rm z} \la 300$ pc), in general our potential matches these
DB models to within 15 per cent out to 4 kpc. We have chosen to
compare our potential to two DB models (Model 1 and Model 4) because
these can be considered to be two extreme cases of the 
models considered in their study, with Model 1 being the least
halo-dominated and Model 4 being the most halo-dominated
(see section 2.7 of \citealt*{BT} for a detailed comparison of these
two models, where Models I and II correspond to DB Models 1 and
4). Over the range where we have kinematic data ($0.5 \la {\rm z/kpc}
\la 2$) it appears that our potential favors Model 1, namely the
model where the disk dominates the circular speed at the solar
neighborhood.

Once we have an estimate for $\Delta\Phi$ we can use this to probe
the vertical mass distribution in the disk through Poisson's
equation ($\partial^2\Phi/\partial{\rm z}^2 = 4\pi G\rho$). Our simple
model corresponds to an infinite razor-thin sheet with a surface mass
density of $a/(2\pi{\rm G}) = 32.5\: \sm {\rm pc}^{-2}$, embedded in
a uniform background with mass density $b/(2\pi{\rm G}) = 0.015\:
\sm {\rm pc}^{-3}$.
Note that this background mass is close to the 
$0.014\:\sm {\rm pc}^{-3}$ predicted using isothermal spherical halo
models \citep[equation 4.279 of][]{BT}. If we assume our background
mass represents the dark halo, it corresponds to a local dark matter
density of 0.57 ${\rm GeV\;cm^{-3}}$, which is noticeably larger than
the canonical value of 0.30 ${\rm GeV\;cm^{-3}}$ typically assumed
\citep[e.g.][]{Ju96}. As pointed out by various authors
\citep[e.g.][]{Ga95,We10,Ga11}, the local dark matter density is
uncertain by a factor of at least 2. Our analysis adds still more
weight to the argument the local halo density may be  substantially
underestimated by the canonical value of 0.30 ${\rm GeV\;cm^{-3}}$,
and this is of immediate interest to dark matter experimentalists.

Perhaps more robust than the local mass density is the surface mass
density. By integrating our mass distribution we obtain a total
surface mass density of $\Sigma_{1.1\rm kpc} = 66 \: \sm {\rm pc}^{-2}$, which 
agrees well with the classical value of $71 \pm 6 \:\sm {\rm pc}^{-2}$
from \citet{Ku91}. If we integrate beyond 1.1 kpc, we find
$\Sigma_{2\rm kpc} = 94 \: \sm {\rm pc}^{-2}$ and 
$\Sigma_{4\rm kpc} = 155 \: \sm {\rm pc}^{-2}$.

Although we would ideally like to quote uncertainties for the above
quantities, this is difficult due to the number of assumptions and
approximations in the adopted prescription. In particular, this method
is dependent on what we assume for the density distributions for our
tracer populations (in our case, the Juri\'c/Ivezi\'c profiles). As a
consequence, the formal errors from the $\chi^2$ fitting will be
meaningless and we choose not to quote them.  We have also
experimented by adding a third term in the expansion of the potential
(equation \ref{eq:pot_expansion}), but found that gave no statistical
improvement to the fit and had no significant influence on the
resulting potential.

\begin{table}
\caption{Best-fit parameters from the potential fitting
  procedure. Parameters $a$ and $b$ correspond to the potential given in
  equation (\ref{eq:pot_expansion}), $\nu(0)$ is the
  solar-neighborhood density normalization and ${\rm z_{h,thin}}$ and
  ${\rm z_{h,thick}}$ are the scale-heights of the thin and thick
  components of the disk profile, respectively. Note that the
  scale-height of the thick component is not a free parameter and is
  kept fixed at three times that of the thin component 
  \citep[following][]{Juric08}.}
\begin{tabular}{lcc}
\hline
Parameter & $\feh$ & Value\\
& Range &\\
\hline
$a\:({\rm km^2\,s^{-2}\,kpc^{-1}})$ & $-$ & -876.7\\
$b\:({\rm km^2\,s^{-2}\,kpc^{-2}})$ & $-$ & -414.4\\\\
$\nu(0)\:(10^{-3}\sm\,{\rm pc}^{-3})$ & $(-0.5,+0.2)$ & 6.1\\
${\rm z_{h,thin}}$ (kpc) & '' & 0.05\\
${\rm z_{h,thick}}$ (kpc) & '' & 0.16\\
$\dispvr(0)\:(\kms)$ & '' & 14.9\\\\
$\nu(0)\:(10^{-3}\sm\,{\rm pc}^{-3})$ & $(-0.8,-0.5)$& 2.7\\
${\rm z_{h,thin}}$ (kpc) & '' & 0.14\\
${\rm z_{h,thick}}$ (kpc) & '' & 0.42\\
$\dispvr(0)\:(\kms)$ & '' & 26.3\\\\
$\nu(0)\:(10^{-3}\sm\,{\rm pc}^{-3})$ & $(-1.5,-0.8)$ & 0.8\\
${\rm z_{h,thin}}$ (kpc) & '' & 0.26\\
${\rm z_{h,thick}}$ (kpc) & '' & 0.78\\
$\dispvr(0)\:(\kms)$ & '' & 38.2\\
\hline
\end{tabular}
\label{tab:potential}
\end{table}

%%%%%%%%%%%%%%%%%%%%%%%%%%%%%%%%%%%%%%%%%%%%%%%%%%%%%%%%%%%%%%%%%%%%%%%%%

\section{Conclusions}
\label{sec:conclusions}

We have undertaken a kinematic study of the Galactic disk using
data from the SDSS equatorial stripe region towards the South
Galactic cap (Stripe 82). By combining spectroscopic data from
\citet{Le08} with proper motions from \citet{Br08}, we have
constructed a sample of 7280 disk stars with full 3-dimensional
positions and velocities, along with $\feh$. This data have allowed us
to probe the kinematics of the disk, tracing means and dispersions as
a function of height from the plane and $\feh$.

These data can be used to investigate the evolution of the disk,
as stars which are born on circular orbits in the plane are heated by
various mechanisms, such as the those caused by spiral arms, the bar,
molecular clouds or accretion events. One avenue for addressing this
is to investigate the ratio $\dispvz/\dispvr$. For the metal-rich
stars we measure this to be 0.6, which is consistent with predictions
for disk heating via spiral arms and molecular clouds
\citep{Je92}. However, predictions from models of satellite accretion
also cover the range we observe and so it is difficult to deduce any
strong conclusions. We also investigate the ratio $\dispvphi/\dispvr$,
and find that in general stars further from the plane exhibit larger
ratios, although this trend is less evident for the more metal-rich
stars. It will be interesting in future for models to
attempt to explain the gradients, both in z and $\feh$, which are
evident in our data. In particular, it is clear that the metal-poor
stars are more isotropic in their kinematics than their metal-rich
counterparts.

Our data also allow us to measure the covariances and constrain the
orientation of the velocity ellipsoid. We found that the
tilt term ($\alpha_{\rm Rz}$) is not consistent with zero, i.e. the
ellipsoid is not aligned with cylindrical polar co-ordinates. Our
results are consistent with previous studies, which have found that
the vertical component of the ellipsoid is close to being aligned in
spherical polar co-ordinates \citep[e.g.][]{Siebert08}. The vertex
deviation ($\alpha_{\rm R\phi}$) is found to be consistent with an
extrapolation to the solar-neighborhood, with a marginal detection
of an increase as one moves away from the plane. If this gradient can
be confirmed it may have interesting implications, as the mechanisms
that are believed to drive this term away from zero (e.g. the bar or
spiral arms) should have greater influence in the plane.

In order to address the nature of disk heating and to disentangle the
contributions from various mechanisms, one needs to go beyond the work
presented here.  The most crucial improvement will be the ability to
make direct estimates for stellar ages, rather than relying on
correlations with metallicity. Ages are notoriously difficult to
measure robustly, even for surveys of bright nearby stars
\citep[e.g.][]{Ho09}. Extending such studies beyond our immediate
solar-neighborhood will be a difficult task, but one that is
currently showing promise \citep[e.g.][]{Bu11}. One aspect that will
help us in this effort is by folding in measurements of alpha-element
abundances that are now being determined routinely for vast numbers of
stars \citep[e.g.][]{Bo11,Le11,Na11,Ru10}. In particular \citet{Bo11}
and \citet{Le11} show what is possible with SDSS data when one
incorporates information on alpha-element abundances.

In recent years there has been some debate regarding the rotation lag
of the disk. The lag has been known about for many years, with recent
determinations of around 30 $\kms{\rm kpc}^{-1}$
\citep[e.g.][]{Gi06}. We find gradients of around 15 to 40 $\kms{\rm
kpc}^{-1}$.
These gradients depend on height from the plane (with the gradients 
in general becoming shallower as z increases) and also on metallicity.
Our findings address one particular bone of contention, which is
whether there exists a correlation between rotation lag and metallicity
\citep[e.g.][]{Iv08,Sp10}. Our data clearly show that the more
metal-poor (and hence hotter) stars exhibit greater lag than their
metal-rich counterparts, with the metal-poor stars reaching a lag of
more than 80 $\kms$ at 2 kpc from the plane.
Our findings are in good qualitative agreement with simulations,
although further work needs to be done to fully exploit these
observational results.

In passing, it is interesting to compare our results on the vertical
gradient of the lag to that for neutral H\,{\small I} gas. For the
Milky Way this has been found to be $15 \pm 4$ $\kms{\rm kpc}^{-1}$
\citep{Ma11}, which is comparable to our findings for metal-rich
stars. This measurement is arguably easier to determine for external
edge-on disk galaxies, where it has been found to be 10 -- 30
$\kms{\rm kpc}^{-1}$ \citep{He07,Ka07,Zs11}. Although the gas response
to the various heating mechanisms will differ from that of the
collisionless stars, the magnitude of the lags appear similar.

Another limitation of our analysis is that we are confined to the 250
square degrees of Stripe 82. Although we are able to probe gradients
with height from the plane, we are unable to draw any conclusions
regarding radial gradients. Furthermore, the limited sky coverage
means that we are susceptible to bias should kinematic substructure be
present within any of our bins. Although no such substructure is
immediately evident in our data, we cannot exclude such an
occurrence. By extending this to the whole SDSS footprint using the
recently released 8th Data Release \citep{Ai11} it will be possible to
overcome some of these limitations, although the lack of radial
coverage will persist as SDSS has very few fields at low Galactic
latitude. Future work that will help to fill in the missing
information at low latitudes includes the LAMOST telescope
\citep[e.g.][]{Wu11} surveys, which will in-part focus on the Galactic
disk. Looking further ahead the field will undoubtedly be
revolutionized by the Gaia satellite mission, which will provide
precise distances and proper motions for billions of stars in our
Galaxy \citep{Pe01,Su09}.

We have also shown an application of our data, employing the Jeans
equations to provide a simple model of the Galactic potential close to
the disk. We use the density model of \citet{Juric08} and our measured
values on the variation of $\dispvz$. Our model does a good job of
reproducing the kinematic data for stars in all three ranges of
metallicity. The model is in excellent agreement with Model 1 from
\citet{DB98b}, indicating a preference for a model where the disk,
rather than the halo, dominates the circular speed at the solar
neighborhood. The main uncertainty in this study is the underlying
profile of the tracer populations, since this is something we cannot
recover from our data due to the complicated selection function of the
SDSS spectroscopic survey. Future unbiased spectroscopic surveys, such
as those undertaken by the LAMOST telescope or the Gaia mission, will
allow much more robust constraints to be made. A knowledge of the
Galactic potential near the disk is clearly great importance.
In particular, it allows us to constrain the local dark matter
density, which is a crucial piece of knowledge when evaluating the
prospects for direct and indirect detection of dark matter
\citep[][and references therein]{Be05}. Despite slow progress since
the classical work of \citet{Ku91}, this field is set to have a great
impact in the coming years.

\section*{Acknowledgments}

The authors wish to thank the following people with whom we have had
useful discussions: G. Gilmore, V. Debattista, V. Belokurov, J. An and
M. Juri\'c. We are also grateful to the anonymous referee for
suggestions which helped improve the clarity of the paper.
The code for calculating the potential of the \citet{DB98b}
models was kindly supplied by W. Dehnen.
MCS acknowledges financial support from the Peking University One
Hundred Talent Fund (985) and NSFC grants 11043005 and 11010022
(International Young Scientist). This work was also supported by the
European Science Foundation (ESF) for the activity entitled 'Gaia
Research for European Astronomy Training'.

Funding for the SDSS and SDSS-II has been provided by the Alfred
P. Sloan Foundation, the Participating Institutions, the National
Science Foundation, the U.S. Department of Energy, the National
Aeronautics and Space Administration, the Japanese Monbukagakusho, the
Max Planck Society, and the Higher Education Funding Council for
England. The SDSS Web Site is http://www.sdss.org/.

\appendix

\section{The turn-off correction for the photometric parallax relation}
\label{app:turn-off}

Our distances are estimated using the photometric parallax relation of
\citet{Iv08}, with one minor modification regarding the
treatment of the main-sequence turn-off. \citet{Iv08} based their
turn-off correction on the color-magnitude sequence for cluster M13,
which has an age of around 10 Gyr and $\feh = -1.54$. Although this is
a sound approach for stars belonging to the Galactic halo, disk stars
will be (in general) both younger and more metal-rich than this
cluster. Therefore we adopt an alternative approach to modelling the
turn-off correction.

We follow the approach described in \citet{Sm09}, where stellar
models were used to construct a metallicity-dependent turn-off
correction. Since we cannot discriminate between different populations
on a stars-by-star basis, we must construct a global correction which
reflects the relative numbers of thin-disk, thick-disk and halo stars
at a given $\feh$. This is done by constructing a toy model to
represent the properties of our sample, with parameters given in Table
\ref{tab:turn_off}. 

The final ingredient for this calculation are theoretical stellar
models, which we take from \citet{Do08}. We downloaded three sets of
models, corresponding to each of our three Galactic components. The
prescription we adopt is the same as \citet{Sm09}. In brief we
we shift each sequence so that it has $M_r=0$ for
$(g\!-\!i)=0.6$ and then, for a given {\feh}, we calculate a weighted
mean $M_r$ as a function of $(g\!-\!i)$ in the range
$0.3<(g\!-\!i)<0.6$, considering only model data up to the
main-sequence turn-off. The weights are determined from the above toy
model, so that we account for the contribution of each Galactic
component at a given metallicity.

\begin{figure}
\plotme{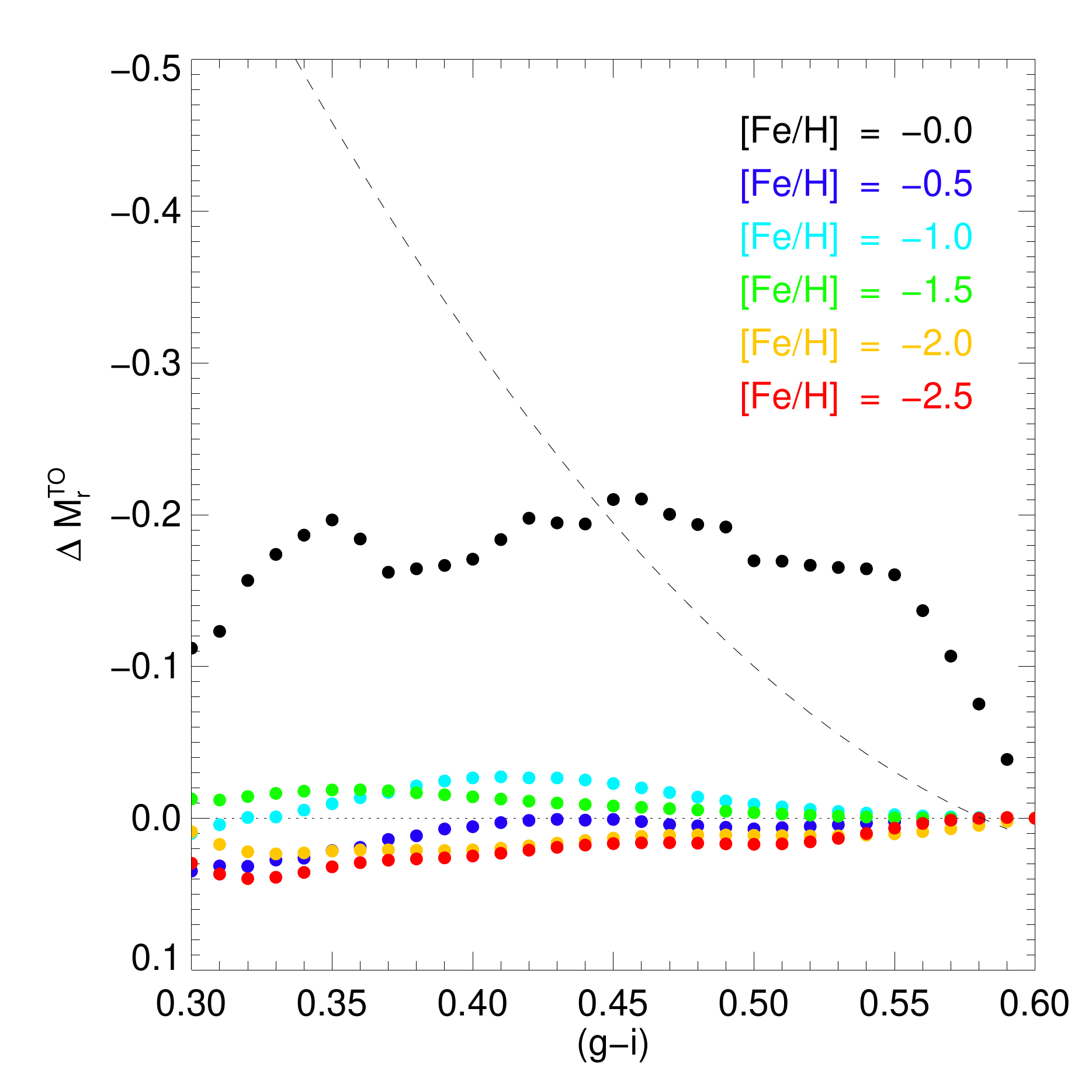}
\caption{Our correction to the photometric parallax relation to
account for main-sequence turn-off stars. The points correspond to the
difference in magnitude between the stellar models and the parallax
relation given in equations (A1$-$A5) of \citet{Iv08}. The dashed line
denotes the turn-off correction given by \citet{Iv08}, which is based
on the halo globular cluster M13.}
\label{fig:turn_off}
\end{figure}

We then calculate the offset between the weighted mean model magnitude
and the uncorrected relation of \citet{Iv08}, i.e. given in equations
(A1$-$A5). This is shown in Fig. \ref{fig:turn_off}. Clearly, for
metallicities below solar all of the models lie around
$\Delta {\rm M_r^{TO}}=0$, i.e. they agree with the $uncorrected$
parallax relation of \citet{Iv08}. Therefore we decide not to
incorporate any turn-off correction when calculating our distances.

The only exception from this behavior is the solar-metallicity model,
which is systematically offset by between 0.1 and 0.2 magnitudes. We
believe this is due to problems with the normalization of the models
(which we do by forcing all models to pass through the point ${\rm
M_r}=0$ at $(g\!-\!i)=0.6$). In any case, this should have little
effect on our results as very few metal-rich stars are blue enough to
lie in this turn-off region; only one per cent of our sample have
$\feh>-0.25$ and $(g\!-\!i)<0.6$.

There will be additional uncertainties on the parallax relation in
this turn-off regime due to scatter in this correction. To account for
this we calculate the standard deviation in ${\rm M_r}$ when we calculate
the mean. The scatter varies as a function of color and metallicity,
but if we take the relation with the largest scatter (corresponding to
$\feh=-1$ dex) we find the following relation, 
\beq
\delta\left(\Delta {\rm M_r^{TO}}\right) = 0.272 - 0.454\,(g\!-\!i).
\eeq
When estimating distances for our stars we
add this uncertainty in quadrature to the other sources of error for
stars in this color range. 

\begin{table}
\caption{Parameters of the toy model used to construct the turn-off
correction to the photometric parallax relation. The parameters are
chosen to be consistent with the following references: densities from
\citet{Juric08}, which we evaluate at $z=1$ kpc, i.e. the mean value
of our sample; $\feh$ from \citet{So03} for the disk and \citet{Iv08}
for the halo; $[\alpha/{\rm Fe}]$ from \citet{Ve04}; ages from
\citet{Nordstrom04} for the disk and chosen to match the halo age
used in \citep{Sm09}.}
\begin{tabular}{lcccccc}
\hline
Component & \multicolumn{2}{c}{Age} & \multicolumn{2}{c}{$\feh$} &
$[\alpha/{\rm Fe}]$ & $\nu\left( z=-1\,{\rm kpc} \right)$\\
 & \multicolumn{2}{c}{(Gyr)} & \multicolumn{2}{c}{(dex)} & (dex) & ($\sm.{\rm pc}^{-3}$)\\
& Mean & Sigma & Mean & Sigma & & \\
\hline
Thin Disk & 2 & 1 & -0.17 & 0.26 & 0.0 & 0.036\\
Thick Disk & 8 & 4 & -0.48 & 0.32 & 0.2 & 0.040\\ 
Halo & 10 & 2 & -1.46 & 0.3 & 0.3 & 0.005\\
\hline
\end{tabular}
\label{tab:turn_off}
\end{table}

\clearpage

\setcounter{table}{0}
\tabletypesize{\tiny}
\def\arraystretch{2.0}
\begin{table}
\caption{Kinematic properties of the disc. The values of
$\covrz$ and $\alpha_{\rm Rz}$ for stars with $\feh < -0.8$ dex
are unreliable due to the difficulty in making the halo correction and
hence have been omitted from this table. Note that $\covrz$ and
$\alpha_{\rm Rz}$ were calculated using only stars within 7.5 kpc $<$ R
$<$ 8.5 kpc and so the number used is less than that quoted in the second
column, especially at large z.}
\begin{tabular}{cccccccccccccccc}
\hline
$\feh$ & $\langle {\rm z} \rangle$ & No. & Halo & $\langle \vR \rangle$ & $\langle \vphi \rangle$ &  $\langle \vz \rangle$ & $\dispvr$ & $\dispvphi$ & $\dispvz$ & $\covrz$ & $\alpha_{\rm Rz}$ & $\covrphi$ & $\alpha_{\rm R\phi}$ \\
(dex) & (kpc) & stars & Frac. & ($\kms$) & ($\kms$) & ($\kms$) & ($\kms$) &
($\kms$) & ($\kms$) & (${\rm km^2 s^{-2}}$) & ($^\circ$) & (${\rm km^2 s^{-2}}$) & ($^\circ$) \\
\hline
$\left(-0.5,+0.2\right)$&$-0.59$&736&$0.00$&$-0.0^{+1.6}_{-1.6}$&$-196.9^{+1.1}_{-1.1}$&$-2.7^{+0.9}_{-1.0}$&$38.8^{+1.2}_{-1.3}$&$27.7^{+0.4}_{-0.4}$&$22.4^{+0.7}_{-0.7}$&$-99.5^{+36.8}_{-36.8}$&$-5.6^{+2.1}_{-2.1}$&$110.4^{+45.6}_{-44.7}$&$8.3^{+3.4}_{-3.4}$\\[3pt]
$\left(-0.5,+0.2\right)$&$-0.86$&736&$0.00$&$3.9^{+1.9}_{-1.8}$&$-188.4^{+1.2}_{-1.2}$&$-1.9^{+1.1}_{-1.1}$&$42.6^{+1.5}_{-1.4}$&$29.8^{+0.1}_{-0.1}$&$25.7^{+0.9}_{-0.9}$&$-322.5^{+66.6}_{-66.7}$&$-14.6^{+2.9}_{-2.8}$&$206.9^{+60.9}_{-61.5}$&$12.0^{+3.5}_{-3.4}$\\[3pt]
$\left(-0.5,+0.2\right)$&$-1.11$&736&$0.00$&$5.6^{+1.9}_{-1.9}$&$-183.9^{+1.4}_{-1.4}$&$-6.0^{+1.2}_{-1.3}$&$43.5^{+1.6}_{-1.6}$&$33.6^{+1.1}_{-1.1}$&$28.5^{+1.0}_{-1.0}$&$-242.3^{+85.6}_{-86.5}$&$-12.0^{+4.1}_{-4.1}$&$100.3^{+76.8}_{-76.5}$&$7.4^{+5.5}_{-5.6}$\\[3pt]
$\left(-0.5,+0.2\right)$&$-1.53$&736&$0.02$&$0.8^{+2.4}_{-2.4}$&$-179.3^{+1.6}_{-1.6}$&$-5.8^{+1.5}_{-1.4}$&$50.2^{+2.1}_{-2.2}$&$35.6^{+1.0}_{-0.9}$&$30.7^{+1.2}_{-1.2}$&$-362.9^{+104.2}_{-106.0}$&$-12.4^{+3.5}_{-3.5}$&$420.1^{+126.5}_{-127.1}$&$16.8^{+4.6}_{-4.6}$\\[13pt]
$\left(-0.8,-0.5\right)$&$-0.59$&543&$0.00$&$0.5^{+2.0}_{-2.0}$&$-190.8^{+1.4}_{-1.5}$&$0.0^{+1.5}_{-1.5}$&$42.7^{+1.5}_{-1.6}$&$32.8^{+1.0}_{-1.0}$&$31.8^{+1.1}_{-1.1}$&$-193.6^{+71.9}_{-70.9}$&$-12.7^{+4.6}_{-4.6}$&$137.1^{+76.4}_{-75.3}$&$10.0^{+5.6}_{-5.5}$\\[3pt]
$\left(-0.8,-0.5\right)$&$-0.96$&543&$0.00$&$6.1^{+2.4}_{-2.3}$&$-174.2^{+1.8}_{-1.8}$&$-1.8^{+1.8}_{-1.8}$&$48.2^{+1.9}_{-1.8}$&$39.6^{+1.5}_{-1.5}$&$38.8^{+1.4}_{-1.4}$&$-154.0^{+127.6}_{-127.8}$&$-10.2^{+8.3}_{-8.4}$&$-89.0^{+110.7}_{-110.3}$&$-6.6^{+8.3}_{-8.3}$\\[3pt]
$\left(-0.8,-0.5\right)$&$-1.26$&543&$0.01$&$4.3^{+2.6}_{-2.5}$&$-168.6^{+2.1}_{-2.1}$&$-3.1^{+1.9}_{-1.9}$&$50.3^{+2.1}_{-2.1}$&$44.6^{+1.9}_{-1.9}$&$37.2^{+1.5}_{-1.5}$&$-181.0^{+133.7}_{-135.0}$&$-8.7^{+6.3}_{-6.4}$&$143.9^{+144.6}_{-144.0}$&$13.5^{+13.5}_{-13.5}$\\[3pt]
$\left(-0.8,-0.5\right)$&$-1.69$&543&$0.04$&$4.0^{+3.2}_{-3.2}$&$-162.1^{+2.4}_{-2.4}$&$-5.0^{+2.2}_{-2.2}$&$57.6^{+2.7}_{-2.7}$&$44.2^{+2.4}_{-2.5}$&$39.6^{+1.8}_{-1.8}$&$-311.3^{+179.3}_{-179.8}$&$-9.7^{+5.6}_{-5.5}$&$519.6^{+227.6}_{-225.0}$&$18.3^{+7.3}_{-7.1}$\\[13pt]
$\left(-1.5,-0.8\right)$&$-0.60$&541&$0.11$&$6.1^{+3.0}_{-2.9}$&$-177.5^{+2.1}_{-2.1}$&$0.6^{+2.2}_{-2.1}$&$52.9^{+2.7}_{-2.7}$&$40.8^{+2.0}_{-2.0}$&$40.2^{+1.8}_{-1.8}$&---&---&$133.6^{+418.9}_{-406.6}$&$5.2^{+17.7}_{-18.7}$\\[3pt]
$\left(-1.5,-0.8\right)$&$-0.99$&541&$0.14$&$4.3^{+3.2}_{-3.3}$&$-160.6^{+2.6}_{-2.6}$&$0.1^{+2.6}_{-2.6}$&$53.0^{+3.2}_{-3.2}$&$42.6^{+2.6}_{-2.7}$&$46.1^{+2.1}_{-2.1}$&---&---&$-275.6^{+473.9}_{-471.1}$&$-10.7^{+21.9}_{-19.3}$\\[3pt]
$\left(-1.5,-0.8\right)$&$-1.32$&541&$0.17$&$5.1^{+3.9}_{-4.0}$&$-150.4^{+3.3}_{-3.3}$&$1.0^{+2.8}_{-2.8}$&$61.3^{+4.1}_{-4.1}$&$50.7^{+2.7}_{-2.7}$&$45.7^{+2.4}_{-2.4}$&---&---&$359.8^{+555.2}_{-551.9}$&$11.8^{+18.9}_{-21.0}$\\[3pt]
$\left(-1.5,-0.8\right)$&$-1.73$&541&$0.27$&$7.6^{+4.6}_{-4.7}$&$-134.5^{+4.0}_{-4.0}$&$1.2^{+3.4}_{-3.3}$&$60.2^{+4.6}_{-4.6}$&$55.3^{+3.0}_{-3.0}$&$45.4^{+3.0}_{-3.0}$&---&---&$-171.3^{+496.2}_{-492.8}$&$-8.7^{+36.8}_{-32.1}$\\[3pt]
\hline
\end{tabular}
\label{tab:kinematics}
\end{table}

\end{document}